\documentclass[usenatbib,a4paper]{mn2e}

\usepackage{graphics}
\usepackage{graphicx}
\usepackage{natbib}
\usepackage{paralist}
\usepackage{subfigure}
\usepackage{amssymb}
\usepackage{amsmath}	
\usepackage{color}

\title[X-ray and neutral hydrogen absorption in radio AGN]
{Connecting X-ray absorption and 21\,cm neutral hydrogen absorption in obscured radio AGN}
\author[V.A. Moss et al.]
{\parbox{\textwidth}{V.A.~Moss,$^{1,2}$\thanks{E-mail: \texttt{vanessa.moss@sydney.edu.au}}
J.R.~Allison$^{3}$,
E.M.~Sadler$^{1,2}$,
R.~Urquhart$^{4}$,
R.~Soria$^{4,1,5}$,
J.R.~Callingham$^{6}$,
S.J.~Curran$^{7}$,
A.~Musaeva$^{1,2,3}$,
E.K.~Mahony$^{1,2}$,
M.~Glowacki$^{1,2,3}$,
S.A.~Farrell$^{1,2}$,
K.W.~Bannister$^{3}$,
A.P.~Chippendale$^{3}$,
P.G.~Edwards$^{3}$,
L.~Harvey-Smith$^{3}$,
I.~Heywood$^{3,8}$,
A.W.~Hotan$^{3}$,
B.T.~Indermuehle$^{3}$,
E.~Lenc$^{1,2}$,
J.~Marvil$^{3}$,
D.~McConnell$^{3}$,
J.E.~Reynolds$^{3}$,
M.A.~Voronkov$^{3}$,
R.M.~Wark$^{3}$,
M.T.~Whiting$^{3}$
}\vspace{0.4cm}\\
\\
\parbox{\textwidth}{$^1$ Sydney Institute for Astronomy, School of Physics A2, The University of Sydney, NSW 2006, Australia \\
$^2$ ARC Centre of Excellence for All-Sky Astrophysics (CAASTRO)\\
$^3$ CSIRO Astronomy and Space Science, ATNF, PO~Box~76, Epping NSW~1710, Australia.\\
$^4$ International Centre for Radio Astronomy Research, Curtin University, GPO Box U1987, Perth, WA 6845, Australia\\
$^5$ National Astronomical Observatories, Chinese Academy of Sciences, Beijing 100012, China\\
$^6$ ASTRON, the Netherlands Institute for Radio Astronomy, PO Box 2, 7990 AA, Dwingeloo, The Netherlands\\
$^7$ School of Chemical and Physical Sciences, Victoria University of Wellington, PO Box 600, Wellington 6140, New Zealand\\
$^8$ Department of Physics and Electronics, Rhodes University, PO Box 94, Grahamstown 6140, South Africa
}}

\let\leqslant=\leq 
\date{}
\def\adet{20 }
\def\anon{74 }
\def\atot{94 }

\begin{document}
\maketitle
\newcommand{\setthebls}{
}
\setthebls
\begin{abstract}
Many radio galaxies show the presence of dense and dusty gas near the active nucleus. This can be traced by both 21\,cm H{\sc i} absorption and soft X-ray absorption, offering new insight into the physical nature of the circumnuclear medium of these distant galaxies. To better understand this relationship, we investigate soft X-ray absorption as an indicator for the detection of associated H{\sc i} absorption, as part of preparation for the First Large Absorption Survey in H{\sc i} (FLASH) to be undertaken with the Australian Square Kilometre Array Pathfinder (ASKAP). We present the results of our pilot study using the Boolardy Engineering Test Array, a precursor to ASKAP, to search for new absorption detections in radio sources brighter than 1\,Jy that also feature soft X-ray absorption. Based on this pilot survey, we detected H{\sc i} absorption towards the radio source PKS\,1657$-$298 at a redshift of $z$ = 0.42. This source also features the highest X-ray absorption ratio of our pilot sample by a factor of 3, which is consistent with our general findings that X-ray absorption predicates the presence of dense neutral gas. By comparing the X-ray properties of AGN with and without detection of H{\sc i} absorption at radio wavelengths, we find that X-ray hardness ratio and H{\sc i} absorption optical depth are correlated at a statistical significance of 4.71$\sigma$. We conclude by considering the impact of these findings on future radio and X-ray absorption studies.
\end{abstract}


\section{Introduction}
The cores of active galaxies produce emission across the electromagnetic spectrum, with each wavelength opening a different window into the nature of the supermassive black holes at their centre and their surrounding medium \citep[][and references therein]{Heckman:2014gc,Tadhunter:2016br}. Many compact active galactic nuclei (AGN) detected at radio wavelengths are also X-ray bright, and the X-ray emission observed can be produced via a number of different mechanisms, such as the inner accretion flow (either an advection-dominated region, or a hot corona above the inner disk and its reflected component), directly from the jet (synchrotron emission), or from the interaction of the jet with the surrounding medium (hot thermal-plasma emission), among other processes \citep[see e.g.][]{Haardt:1993hb,Merloni:2003kv,Turner:2009ik,Worrall:2009jm,Fabian:2012gi,Reynolds:2016dr}.

Previous studies investigating X-ray and radio AGN have included comparison of the column densities estimated from 21\,cm neutral atomic hydrogen (H{\sc i}) absorption ($N_{\rm HI}$, assuming a spin temperature) and an absorbed X-ray spectrum ($N_{\rm H}$, the combination of ionised, neutral and molecular hydrogen), finding a weak correlation between these two properties \citep[e.g.][]{Vink:2006bh,Ostorero:2010eb}. This in turn indicates likely spatial correlation between these different wavelength tracers of gas near the AGN, although their exact relationship is of ongoing consideration as the X-ray emission most likely originates close to the central black hole while the radio emission can be more extended \citep{Vink:2006bh}. A specific focus has been on the gigahertz-peaked spectrum (GPS) and compact steep spectrum (CSS) radio sources, as these are believed to represent the early stages of AGN evolution \citep{1995A&A...302..317F,ODea:1998jt,2006A&A...445..889T,Siemiginowska:2008ik}. This has been supplemented by studies showing a higher detection rate of H{\sc i} in these kinds of galaxies, particularly in the local Universe \citep[e.g.][]{Gereb:2015cx,Maccagni:2017vn}. However, different processes may dominate in particular AGN depending on their accretion rates, jet power, density of the surrounding medium and, importantly, evolutionary age.  

By investigating the connection between the X-ray properties, radio emission and H{\sc i} absorption, we can build up a better understanding of how these various factors influence the resulting observed properties and consequent evolution of galaxies. Thus, our study extends beyond the youngest radio AGN to the correlation between X-rays and radio in AGN of all types. We are particularly interested in tracing the gas of galaxies, as gas content across cosmic timescales is a key factor in the question of overall evolution of the baryonic Universe. Results show that the gas reservoir has decreased to present day \citep{Noterdaeme:2012fm,Curran:2017tj}, with the overall state of galaxies becoming more quiescent alongside a decrease in the star formation rate \citep{Madau:2014gt}. H{\sc i} absorption studies offer a unique glimpse into these gaseous galaxies along a sightline to the often-compact background radio emission, revealing gas that is `censored' by the size of the radio source. The gas illuminated by these studies is generally found to trace circumnuclear disks, but can also reveal outflow and infall processes \citep[][and references therein]{Morganti:2015vl}. A key advantage is the ability to detect neutral hydrogen gas independent of redshift, in contrast to studies of H{\sc i} emission. 

\begin{figure}
\includegraphics[angle=0,width=0.46\textwidth]{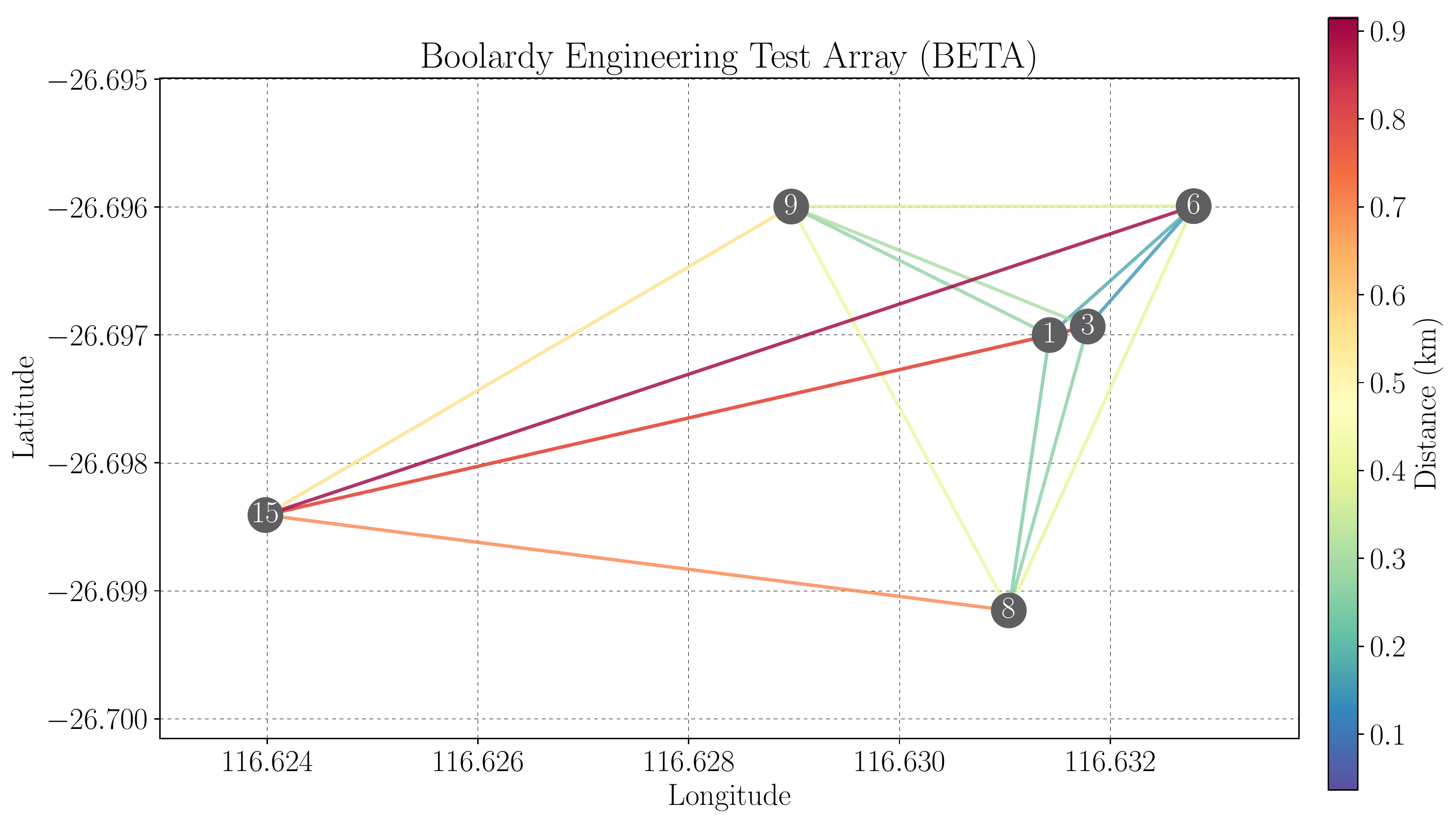}
\caption{Baseline map of the Boolardy Engineering Test Array (BETA), a six-antenna precursor interferometer for the Australian Square Kilometre Array Pathfinder (ASKAP) which was in operation until February 2016. The baseline lengths are colour-coded by distance, ranging from 37--915\,m. For more details about the BETA instrument and baseline properties, see \citet{Hotan:2014dv}.}
\label{fig:beta}
\end{figure}

The First Large Absorption Survey in H{\sc i} (FLASH, Sadler et al. in prep) will be the first blind southern all-sky survey for H{\sc i} absorption. It will be performed by the Australian Square Kilometre Array Pathfinder \citep[ASKAP,][]{Johnston:2007ku}, which features instantaneous 300\,MHz bandwidth and a 5 $\times$ 5 deg field of view. ASKAP will be able to efficiently search the redshift range 0.4 $< z <$ 1.0 for 21\,cm H{\sc i} absorption towards over 150,000 background galaxies, in a single frequency band covering frequencies of 700-1000\,MHz with little impact from radio frequency interference (RFI) due to its radio-quiet host site at the Murchison Radio-astronomy Observatory. Until February 2016, the ASKAP instrument operated in commissioning mode as a six-antenna prototype interferometer known as the Boolardy Engineering Test Array \citep[BETA,][]{Hotan:2014dv,Mcconnell:2016fu}. A map of BETA baselines, including the numbering of relevant antennas, is shown in Figure \ref{fig:beta}. Despite its modest sensitivity, BETA has already highlighted the scope for exploration of H{\sc i} in this unique parameter space with the discovery of associated H{\sc i} absorption towards PKS\,1740$-$517 at $z$ = 0.44 \citep{2015MNRAS.453.1249A,Allison:2016ds}. PKS\,1740$-$517 presented a particularly interesting case study given its diverse multi-wavelength properties. It features a hard X-ray spectrum that is absorbed at the soft end ($<$ 2\,keV), suggesting the presence of dense hydrogen along the line of sight to the source of X-ray emission. We focus here on further investigating this connection between H{\sc i} absorption and soft X-ray absorption in radio galaxies. 

\begin{figure}
\includegraphics[angle=0,width=0.46\textwidth]{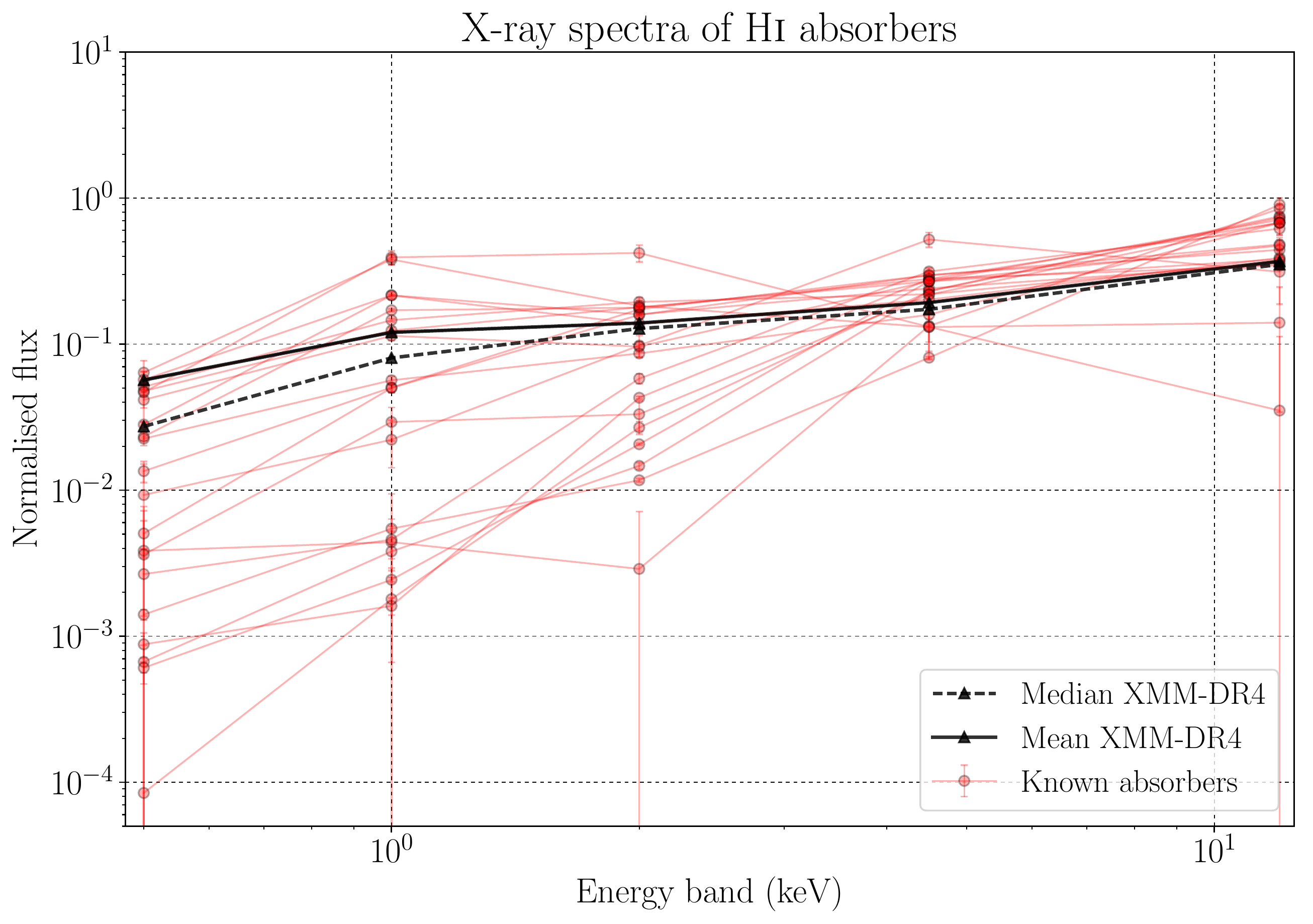}
\includegraphics[angle=0,width=0.46\textwidth]{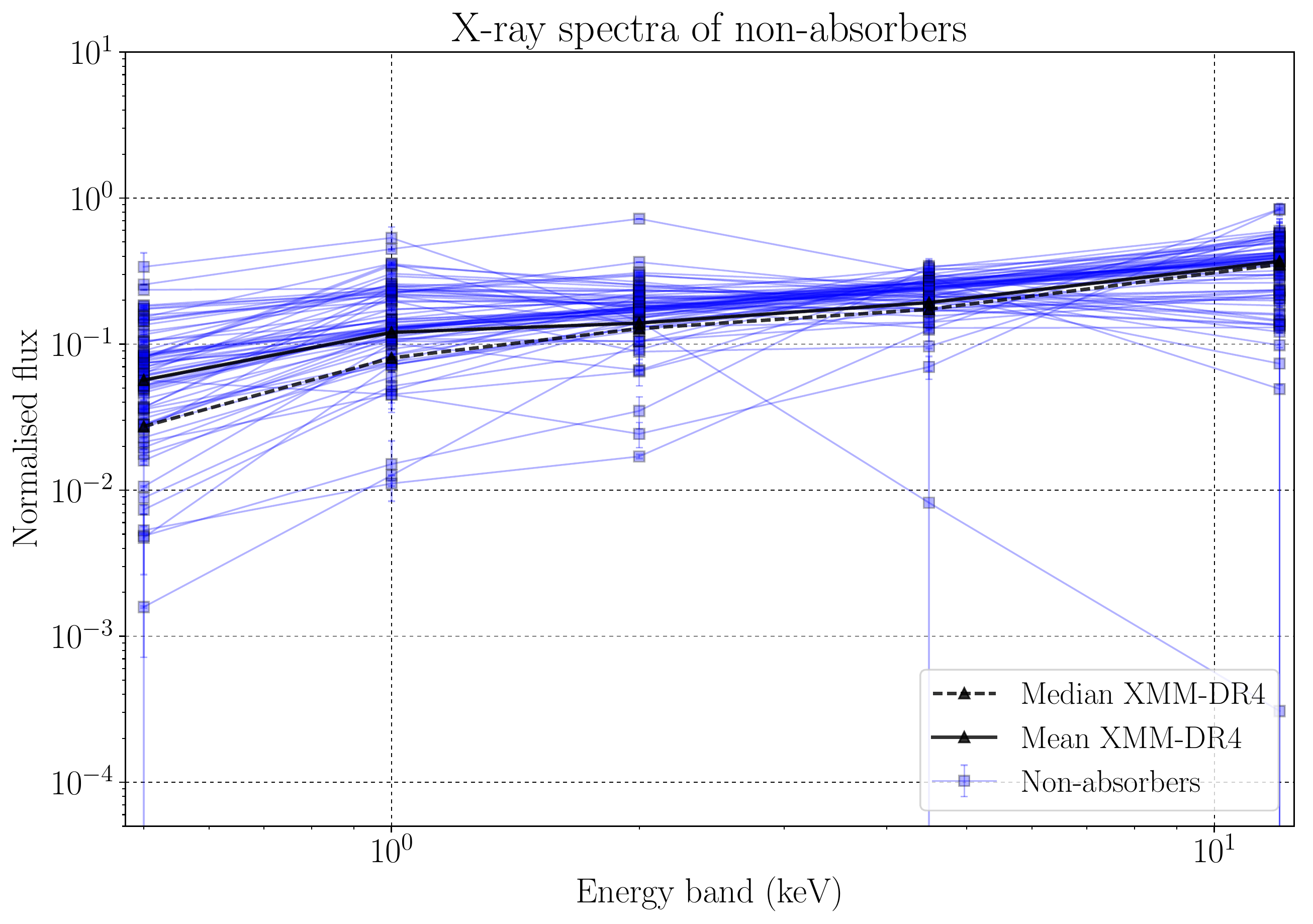}
\caption{3XMM-DR4 spectral properties for all galaxies with known H{\sc i} detections (top, red) and those with no detection of H{\sc i} absorption (bottom, blue), shown on the same axes. The fluxes have been normalised by the catalogued flux for the whole band ($f_8$, 0.2--12\,keV) to minimise the effect of scatter. The population of H{\sc i} absorbers generally features absorbed spectra (indicated by the missing soft X-ray flux at lower energies) and less scatter in flux prior to normalisation. Conversely, the non-absorbers feature flatter spectra and a much larger scatter in flux distribution before normalising. We show the median and mean spectra for all catalogued 3XMM-DR4 sources in black.}
\label{fig:known}
\end{figure}

We consider in this paper the X-ray properties of PKS\,1740$-$517 alongside other known H{\sc i} absorbing sources in associated systems. We compare the properties of these systems to galaxies with non-detections of H{\sc i} absorption that also feature X-ray emission. We also report on the results of the pilot study we have carried out using BETA to target radio-bright, absorbed X-ray sources. We report on our new detection of H{\sc i} absorption towards PKS\,1657$-$298 and discuss the multi-wavelength properties of this radio source. Finally, we summarise our findings, and speculate on the future potential for using X-ray spectral information in the context of informing FLASH and other searches for H{\sc i} absorption. 

\section{X-ray properties of AGN with known H{\sc i} absorption and H{\sc i} non-detections}\label{propcompare}
To examine the connection between H{\sc i} absorption and the properties of associated X-ray emission, we considered both sources with known H{\sc i} absorption and sources which had been searched for H{\sc i} absorption with no resulting detection. We refer throughout the paper to sources without a detection as either non-detections or non-absorbers, but stress that they are only non-absorbing at the limit of current observations. Our list of associated systems that have been searched for H{\sc i} absorption was adapted from a compilation maintained from the literature \citep[see][for an earlier version]{2010ApJ...712..303C}. We note that we do not consider intervening systems here, as we are interested in the properties of neutral gas in radio AGN hosts and so the H{\sc i} should be associated with the AGN rather than along the line-of-sight. We also exclude the BETA sample described later in this paper, given the lack of complete redshift information. We cross-matched the list of absorbers and non-detections against the 3XMM-DR4 catalogue \citep{2015ASPC..495..319R}, resulting in a total of \adet H{\sc i} absorbers and \anon non-H{\sc i} absorbers with corresponding X-ray emission. A list of these \atot sources and their properties is given in Table \ref{tb:sources}. The median separation between the radio and X-ray sources was 0.4$''$, which is well within the average positional offsets in 3XMM-DR4 of $\sim$1-2$''$ when compared with verified SDSS counterparts \citep{2015ASPC..495..319R}.

\begin{figure}
\includegraphics[angle=0,width=0.46\textwidth]{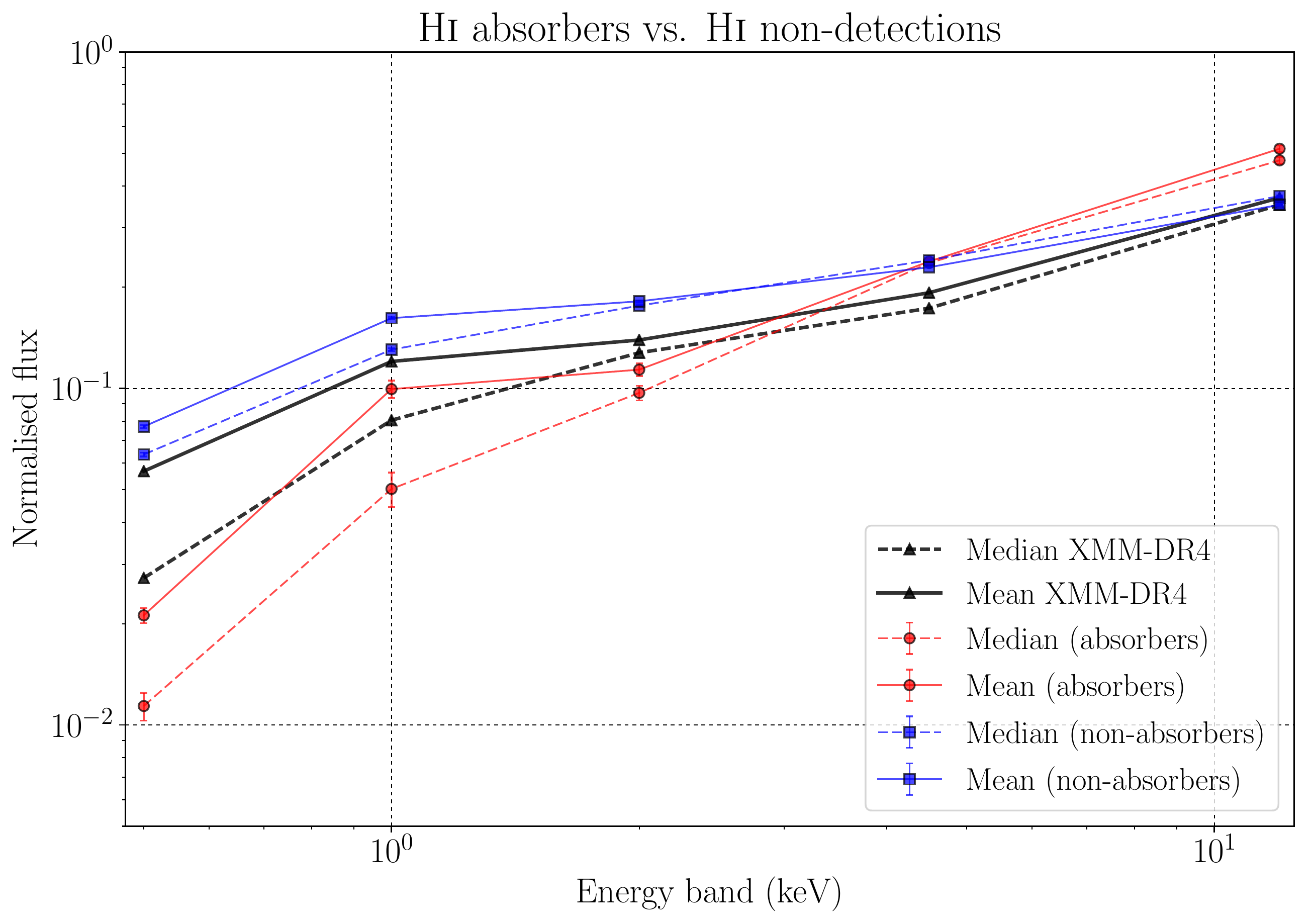}
\caption{The median and mean spectra for H{\sc i} absorbers (red) and H{\sc i} non-absorbers (blue) compared with the overall distribution of all X-ray sources (black). As in the previous figure, fluxes have been normalised by the catalogued flux for the whole band ($f_8$, 0.2--12\,keV) to minimise the effect of scatter. This highlights the more absorbed X-ray spectrum that is common to known H{\sc i} absorbers versus the relative flatness of non-absorber X-ray spectra. The median and mean spectra for all 3XMM-DR4 sources are shown in black.}
\label{fig:average}
\end{figure}

\begin{table*}
{\tiny \centering
\caption{Properties of our sample of 94 AGN with catalogued X-ray emission in the XMM-DR4 catalogue \citep{2015ASPC..495..319R} that either have detected H{\sc i} absorption or have been searched for H{\sc i} absorption with no detection. The overall compilation of H{\sc i} absorbers and non-absorbers was adapted from a list maintained from the literature \citep[see e.g.][]{2010ApJ...712..303C}. We note that $N_{\rm HI}$ is calculated assuming a covering fraction $f$ of 1 and a spin temperature $T_s$ of 100\,K. The variables $f_1$ and $f_5$ refer to the fluxes in Band 1 (0.2--0.5\,keV) and Band 5 (4.5--12\,keV) of 3XMM-DR4 respectively. Uncertainties in values are implicit in the significant figures.}\label{tb:sources}
\begin{tabular}{clcccccrcccr}
\hline
&Source & H{\sc i}& RA & Dec & $z$ & $N_{\rm HI}$& $\tau$ &  Ref. & $f_{1}$ & $f_{5}$& $\frac{f_5}{f_1}$ \\
&&&J2000&J2000&&10$^{20}$\,cm$^{-2}$&&&10$^{-14}$\,mW\,m$^{-2}$&10$^{-14}$\,mW\,m$^{-2}$&\\
\hline
1 & NGC 0315 & Y & 00:57:48.88 & $+$30:21:08.82 & 0.016 & 6.9 & 0.05 & E10 & 3.70 & 46.63 & 13\\ 
2 & CGRaBS J0111+3906 & Y & 01:11:37.32 & $+$39:06:28.12 & 0.668 & 22.9 & 0.44 & O06 & 0.02 & 3.89 & 219\\ 
3 & NGC 1167 & Y & 03:01:42.37 & $+$35:12:20.66 & 0.016 & 1.2 & 0.0023 & E10 & 0.31 & 0.69 & 2\\ 
4 & PKS 0428+20 & Y & 04:31:03.76 & $+$20:37:34.25 & 0.219 & 3.5 & 0.0445 & V03 & 0.04 & 1.23 & 34\\ 
5 & PKS 0500+019 & Y & 05:03:21.20 & $+$02:03:04.68 & 0.585 & 6.2 & 0.045 & C98 & 0.28 & 26.87 & 95\\ 
6 & 2MFGC 06756 & Y & 08:34:54.90 & $+$55:34:21.07 & 0.242 & 1.1 & 0.0032 & V03 & 29.85 & 217.16 & 7\\ 
7 & B2 0902+34 & Y & 09:05:30.10 & $+$34:07:56.93 & 3.398 & 3.1 & 0.011 & U91 & 0.09 & 2.80 & 33\\ 
8 & J125711-172434 & Y & 12:57:11.59 & $-$17:24:33.98 & 0.048 & 33.9 & 0.1 & G17 & 0.65 & 0.49 & 1\\ 
9 & NGC 5090 & Y & 13:21:12.82 & $-$43:42:16.38 & 0.011 & 2.7 & 0.029 & M01 & 2.23 & 33.09 & 15\\ 
10 & 4C +32.44 & Y & 13:26:16.51 & $+$31:54:09.50 & 0.370 & 0.7 & 0.00268 & V03 & 0.64 & 5.15 & 8\\ 
11 & NGC 5363 & Y & 13:56:07.21 & $+$05:15:17.17 & 0.004 & 11.2 & 0.039 & V89 & 1.30 & 8.88 & 7\\ 
12 & 4C +62.22 & Y & 14:00:28.65 & $+$62:10:38.60 & 0.431 & 1.9 & 0.00741 & V03 & 0.12 & 27.95 & 232\\ 
13 & MRK 0668 & Y & 14:07:00.39 & $+$28:27:14.69 & 0.077 & 1.8 & 0.00433 & V03 & 0.93 & 27.94 & 30\\ 
14 & PKS 1549-79 & Y & 15:56:58.87 & $-$79:14:04.27 & 0.150 & 3.6 & 0.026 & M01 & 0.42 & 490.84 & 1167\\ 
15 & 3C 353 & Y & 17:20:28.16 & $-$00:58:46.63 & 0.030 & 41.7 & 0.109 & M01 & 0.07 & 83.78 & 1123\\ 
16 & PKS 1740-517 & Y & 17:44:25.39 & $-$51:44:44.70 & 0.441 & 5.0 & 0.2026 & A15 & 0.01 & 69.30 & 8689\\ 
17 & PKS 1814-63 & Y & 18:19:35.00 & $-$63:45:48.20 & 0.065 & 24.0 & 0.192 & A12 & 0.41 & 313.18 & 770\\ 
18 & 3C 452 & Y & 22:45:48.77 & $+$39:41:16.01 & 0.081 & 6.5 & 0.058 & C13 & 0.37 & 235.38 & 641\\ 
19 & 3C 459 & Y & 23:16:35.23 & $+$04:05:18.06 & 0.220 & 0.7 & 0.00311 & V03 & 0.67 & 7.65 & 11\\ 
20 & COINS J2355+4950 & Y & 23:55:09.46 & $+$49:50:08.34 & 0.238 & 2.8 & 0.018 & V03 & 0.03 & 6.00 & 188\\ 
21 & NGC 0383 & N & 01:07:24.96 & $+$32:24:45.22 & 0.017 & $<$0.6 & $<$0.003 & E10 & 2.36 & 3.94 & 2\\ 
22 & J011132-730209 & N & 01:11:32.25 & $-$73:02:09.89 & 0.067 & $<$7.8 & - & G17 & 0.07 & 9.13 & 122\\ 
23 & UGC 01651 & N & 02:09:38.53 & $+$35:47:51.07 & 0.038 & $<$0.5 & $<$0.003 & E10 & 0.94 & 1.64 & 2\\ 
24 & 3C 066A & N & 02:22:39.61 & $+$43:02:07.80 & 0.444 & $<$0.5 & - & G+ & 36.04 & 89.11 & 2\\ 
25 & UGC 01841 & N & 02:23:11.41 & $+$42:59:31.38 & 0.021 & $<$3.8 & $<$0.0195 & C13 & 4.40 & 9.69 & 2\\ 
26 & 3C 067 & N & 02:24:12.30 & $+$27:50:11.54 & 0.310 & $<$1.4 & $<$0.008 & V03 & 1.28 & 30.28 & 24\\ 
27 & [HB89] 0235+164 & N & 02:38:38.93 & $+$16:36:59.29 & 0.940 & $<$1.3 & - & G+ & 7.66 & 144.75 & 19\\ 
28 & [HB89] 0237-233 & N & 02:40:08.17 & $-$23:09:15.73 & 2.223 & $<$0.2 & - & G+ & 22.30 & 116.81 & 5\\ 
29 & B3 0309+411B & N & 03:13:01.96 & $+$41:20:01.18 & 0.134 & $<$1.7 & $<$0.0087 & C13 & 13.44 & 203.40 & 15\\ 
30 & ESO 248-G006 & N & 03:17:57.67 & $-$44:14:17.45 & 0.076 & $<$0.6 & $<$0.01 & A12 & 233.57 & 918.64 & 4\\ 
31 & NGC 1275 & N & 03:19:48.16 & $+$41:30:42.12 & 0.018 & - & - & G+ & 968.10 & 10082.00 & 10\\ 
32 & UGC 02748 & N & 03:27:54.19 & $+$02:33:41.98 & 0.030 & $<$1.5 & $<$0.028 & M01 & 0.25 & 12.75 & 51\\ 
33 & J033114-524148 & N & 03:31:14.99 & $-$52:41:48.19 & 0.067 & $<$15.8 & - & G17 & 1.38 & 10.00 & 7\\ 
34 & 3C 111 & N & 04:18:21.28 & $+$38:01:35.80 & 0.049 & $<$0.4 & $<$0.002 & C13 & 10.65 & 3865.50 & 363\\ 
35 & J043022-613201 & N & 04:30:22.00 & $-$61:32:01.00 & 0.056 & $<$3.2 & - & G17 & 4.03 & 5.75 & 1\\ 
36 & 3C 120 & N & 04:33:11.10 & $+$05:21:15.62 & 0.033 & $<$0.4 & $<$0.006 & V89 & 222.42 & 3263.90 & 15\\ 
37 & ESO 362-G021 & N & 05:22:57.98 & $-$36:27:30.85 & 0.057 & $<$0.1 & $<$0.013 & V89 & 125.94 & 612.09 & 5\\ 
38 & [HB89] 0537-286 & N & 05:39:54.28 & $-$28:39:55.94 & 3.014 & $<$2.8 & $<$0.1 & C08 & 14.87 & 243.07 & 16\\ 
39 & [HB89] 0552+398 & N & 05:55:30.81 & $+$39:48:49.18 & 2.365 & $<$1.7 & - & G+ & 1.66 & 116.97 & 71\\ 
40 & 3C 153 & N & 06:09:32.51 & $+$48:04:15.38 & 0.277 & $<$0.6 & $<$0.003 & V03 & 0.99 & 0.00 & 0\\ 
41 & PKS 0625-35 & N & 06:27:06.76 & $-$35:29:15.25 & 0.055 & $<$0.4 & $<$0.007 & M01 & 112.89 & 152.73 & 1\\ 
42 & 4C +43.15 & N & 07:35:21.89 & $+$43:44:20.33 & 2.429 & - & $<$0.004 & R99 & 0.10 & 1.53 & 15\\ 
43 & B2 0738+31 & N & 07:41:10.70 & $+$31:12:00.22 & 0.631 & $<$2.4 & - & G+ & 5.72 & 40.14 & 7\\ 
44 & CGCG 262-029 & N & 07:48:36.87 & $+$55:48:58.25 & 0.036 & $<$2.8 & $<$0.0054 & C13 & 0.13 & 0.10 & 1\\ 
45 & NGC 2484 & N & 07:58:28.11 & $+$37:47:11.80 & 0.043 & $<$3.2 & $<$0.0165 & C13 & 3.92 & 9.02 & 2\\ 
46 & 3C 192 & N & 08:05:35.00 & $+$24:09:50.33 & 0.060 & $<$0.4 & $<$0.002 & G06 & 0.60 & 23.44 & 39\\ 
47 & 3C 207 & N & 08:40:47.59 & $+$13:12:23.58 & 0.681 & $<$0.5 & $<$0.003 & V03 & 10.62 & 120.48 & 11\\ 
48 & [HB89] 0836+710 & N & 08:41:24.36 & $+$70:53:42.18 & 2.172 & - & - & G+ & 190.20 & 2687.80 & 14\\ 
49 & OJ +287 & N & 08:54:48.88 & $+$20:06:30.64 & 0.306 & - & - & G+ & 31.00 & 139.49 & 4\\ 
50 & PKS 0921-213 & N & 09:23:38.88 & $-$21:35:47.11 & 0.052 & $<$3.5 & $<$0.059 & A12 & 75.21 & 488.68 & 6\\ 
51 & [HB89] 0954+658 & N & 09:58:47.24 & $+$65:33:54.83 & 0.368 & $<$1.4 & - & G+ & 20.27 & 171.29 & 8\\ 
52 & SDSS J103507.04+562846.8 & N & 10:35:07.04 & $+$56:28:46.81 & 0.459 & $<$1.3 & $<$0.007 & V03 & 0.08 & 4.40 & 55\\ 
53 & [HB89] 1055+018 & N & 10:58:29.61 & $+$01:33:58.82 & 0.888 & $<$0.5 & - & G+ & 34.41 & 162.56 & 5\\ 
54 & MRK 0421 & N & 11:04:27.31 & $+$38:12:31.79 & 0.030 & $<$0.4 & $<$0.006 & V89 & 7841.50 & 4457.70 & 1\\ 
55 & 4C +14.41 & N & 11:20:27.81 & $+$14:20:55.00 & 0.362 & $<$0.6 & $<$0.003 & V03 & 0.19 & 2.01 & 11\\ 
56 & NGC 3665 & N & 11:24:43.67 & $+$38:45:46.26 & 0.007 & $<$20.0 & $<$0.11 & E10 & 6.31 & 9.01 & 1\\ 
57 & NGC 3862 & N & 11:45:05.01 & $+$19:36:22.75 & 0.022 & $<$0.6 & $<$0.008 & V89 & 51.71 & 66.40 & 1\\ 
58 & NGC 4278 & N & 12:20:06.82 & $+$29:16:50.74 & 0.002 & $<$0.3 & $<$0.0015 & E10 & 45.14 & 123.81 & 3\\ 
59 & FBQS J1221+3010 & N & 12:21:21.94 & $+$30:10:37.09 & 0.184 & $<$7.9 & $<$0.044 & G15 & 1449.50 & 1269.30 & 1\\ 
60 & B2 1219+28 & N & 12:21:31.69 & $+$28:13:58.51 & 0.102 & $<$1.2 & - & G+ & 194.43 & 107.34 & 1\\ 
61 & 3C 273 & N & 12:29:06.70 & $+$02:03:08.60 & 0.158 & - & - & G+ & 1465.90 & 6569.30 & 4\\ 
62 & MESSIER 087 & N & 12:30:49.42 & $+$12:23:28.03 & 0.004 & $<$0.7 & $<$0.007 & V89 & 974.97 & 1126.20 & 1\\ 
63 & NGC 4696 & N & 12:48:49.25 & $-$41:18:38.99 & 0.010 & $<$0.5 & $<$0.009 & M01 & 117.31 & 385.18 & 3\\ 
64 & 3C 279 & N & 12:56:11.17 & $-$05:47:21.52 & 0.536 & $<$0.1 & - & G+ & 108.38 & 573.47 & 5\\ 
65 & [HB89] 1308+326 & N & 13:10:28.66 & $+$32:20:43.80 & 0.996 & - & - & G+ & 48.60 & 307.76 & 6\\ 
66 & 3C 287 & N & 13:30:37.69 & $+$25:09:10.87 & 1.055 & $<$0.1 & - & G+ & 8.04 & 26.19 & 3\\ 
67 & IC 4296 & N & 13:36:39.03 & $-$33:57:56.99 & 0.012 & $<$2.2 & $<$0.041 & M01 & 4.01 & 60.66 & 15\\ 
68 & IC 4374 & N & 14:07:29.76 & $-$27:01:04.30 & 0.022 & $<$1.2 & $<$0.017 & V89 & 4.23 & 8.25 & 2\\ 
69 & 3C 303.1 & N & 14:43:14.56 & $+$77:07:27.70 & 0.267 & $<$1.4 & $<$0.008 & V03 & 0.18 & 1.47 & 8\\ 
70 & B3 1445+410 & N & 14:47:12.76 & $+$40:47:44.95 & 0.195 & $<$1.1 & $<$0.006 & G15 & 1.22 & 0.83 & 1\\ 
71 & CGCG 021-063 & N & 15:16:40.22 & $+$00:15:01.91 & 0.052 & $<$0.5 & $<$0.0028 & G06 & 18.67 & 136.99 & 7\\ 
72 & 3C 318 & N & 15:20:05.45 & $+$20:16:05.77 & 1.574 & $<$0.1 & - & G+ & 1.34 & 10.17 & 8\\ 
73 & 2MASX J15320226+3016290 & N & 15:32:02.24 & $+$30:16:28.96 & 0.065 & $<$11.7 & $<$0.064 & G15 & 31.36 & 39.49 & 1\\ 
74 & NGC 6166 & N & 16:28:38.48 & $+$39:33:05.62 & 0.030 & $<$3.7 & $<$0.0192 & C13 & 410.26 & 1299.10 & 3\\ 
75 & NGC 6251 & N & 16:32:31.97 & $+$82:32:16.40 & 0.025 & $<$1.0 & $<$0.013 & V89 & 32.04 & 255.84 & 8\\ 
76 & PKS 1637-77 & N & 16:44:16.12 & $-$77:15:48.82 & 0.043 & $<$2.6 & $<$0.047 & M01 & 13.11 & 102.20 & 8\\ 
77 & MRK 0501 & N & 16:53:52.22 & $+$39:45:36.61 & 0.034 & $<$1.7 & $<$0.023 & V89 & 760.76 & 1051.70 & 1\\ 
78 & B3 1702+457 & N & 17:03:30.38 & $+$45:40:47.17 & 0.060 & $<$4.2 & $<$0.0229 & C11 & 286.77 & 424.09 & 1\\ 
79 & CGCG 173-014 & N & 18:35:03.39 & $+$32:41:46.79 & 0.058 & $<$2.8 & $<$0.0147 & C13 & 558.51 & 2492.80 & 4\\ 
80 & 3C 390.3 & N & 18:42:08.99 & $+$79:46:17.11 & 0.056 & $<$4.0 & $<$0.0189 & C13 & 419.80 & 2784.60 & 7\\ 
81 & PKS 2000-330 & N & 20:03:24.12 & $-$32:51:45.14 & 3.773 & $<$9.5 & - & G+ & 3.76 & 36.83 & 10\\ 
82 & PKS 2008-068 & N & 20:11:14.22 & $-$06:44:03.55 & 0.547 & $<$0.3 & - & G+ & 0.02 & 2.22 & 123\\ 
83 & PKS 2126-15 & N & 21:29:12.18 & $-$15:38:41.03 & 3.268 & $<$3.6 & - & G+ & 30.34 & 656.24 & 22\\ 
84 & PKS 2135-20 & N & 21:37:50.01 & $-$20:42:31.68 & 0.635 & $<$1.1 & $<$0.006 & V03 & 0.32 & 3.12 & 10\\ 
85 & ESO 075-G041 & N & 21:57:05.98 & $-$69:41:23.68 & 0.028 & $<$0.6 & $<$0.01 & M01 & 106.87 & 443.52 & 4\\ 
86 & BL Lac & N & 22:02:43.29 & $+$42:16:39.97 & 0.069 & $<$0.7 & $<$0.009 & V89 & 18.46 & 489.17 & 26\\ 
87 & [HB89] 2201+315 & N & 22:03:14.98 & $+$31:45:38.27 & 0.295 & $<$0.5 & - & G+ & 47.85 & 311.95 & 7\\ 
88 & J220538-053531 & N & 22:05:38.59 & $-$05:35:31.88 & 0.057 & $<$9.8 & - & G17 & 37.20 & 76.99 & 2\\ 
89 & 3C 445 & N & 22:23:49.53 & $-$02:06:12.85 & 0.056 & $<$2.4 & $<$0.008 & M01 & 4.32 & 684.11 & 158\\ 
90 & 3C 454 & N & 22:51:34.74 & $+$18:48:40.10 & 1.757 & $<$0.2 & - & G+ & 2.07 & 11.89 & 6\\ 
91 & 3C 454.3 & N & 22:53:57.75 & $+$16:08:53.56 & 0.859 & $<$0.1 & - & G+ & 61.42 & 902.88 & 15\\ 
92 & IC 1459 & N & 22:57:10.61 & $-$36:27:43.99 & 0.006 & $<$0.7 & $<$0.009 & V89 & 7.77 & 47.53 & 6\\ 
93 & MCG -07-47-031 & N & 23:19:05.91 & $-$42:06:48.20 & 0.055 & $<$2.2 & $<$0.038 & A12 & 68.45 & 89.42 & 1\\ 
94 & NGC 7720 & N & 23:38:29.39 & $+$27:01:53.26 & 0.030 & $<$3.2 & $<$0.0162 & C13 & 1.10 & 2.19 & 2\\ 
\hline
\end{tabular}}
{~\\References: V89 \citep{1989AJ.....97..708V}, U91 \citep{1991PhRvL..67.3328U}, C98 \citep{1998ApJ...494..175C}, R99 \citep{1999mdrg.conf..113R}, M01 \citep{2001MNRAS.323..331M}, V03 \citep{2003A&A...404..861V}, G06 \citep{2006MNRAS.373..972G}, O06 \citep{2006A&A...457..531O}, C08 \citep{2008MNRAS.391..765C}, E10 \citep{2010MNRAS.406..987E}, C11 \citep{2011MNRAS.418.1787C}, A12 \citep{Allison:2012eb}, C13 \citep{2013MNRAS.429.2380C}, G15 \citep{Gereb:2015cx}, G+ (Grasha et al, in prep), G17 \citep{Glowacki:2017cm}.}\\
\end{table*}

\subsection{Results of population comparison}
In order to characterise the X-ray spectral energy distribution, we plotted the spectral shape of both populations as given by the 3XMM-DR4 catalogue in each of the five bands of {\it XMM-Newton} (0.2--0.5\,keV, 0.5--1.0\,keV, 1.0--2.0\,keV, 2.0--4.5\,keV and 4.5--12\,keV, corresponding to bands $f_1$ to $f_5$). Each spectrum is normalised by its catalogued total flux ($f_8$, 0.2--12\,keV), in order to aid comparison and reduce scatter. In this paper, we define X-ray hardness ratio as the ratio of the 4.5--12\,keV band over the 0.2--0.5\,keV band, specified by $\frac{f_5}{f_1}$. The results are shown in Figure \ref{fig:known}, for both the H{\sc i} absorbers (top) and the H{\sc i} non-detections (bottom), with the normalised fluxes plotted at the position of the upper limit of each band. The median and mean spectra for both samples are shown in Figure \ref{fig:average}, compared with the median and mean spectrum for the overall 3XMM-DR4 source population in black,  to highlight the difference in the two populations.. The population of H{\sc i} absorbers generally features absorbed spectra (indicated by the missing soft X-ray flux at the low energy end) and less scatter in flux prior to normalisation. Conversely, the non-absorbers feature flatter spectra and a much larger scatter in flux distribution before normalising. The effect of redshift has not been included in the plotted figures.  Including this effect would shift the spectra left towards lower energies and affect the flux scale, but is most relevant only for sources with $z$ $>$ 1 (a small fraction of our sample). Under the assumption that we are dealing with sources featuring simple power laws, which is true in the majority of cases, then this does not have a significant impact on the measured ratio of $\frac{f_5}{f_1}$.

\begin{figure}
\includegraphics[angle=0,width=0.46\textwidth]{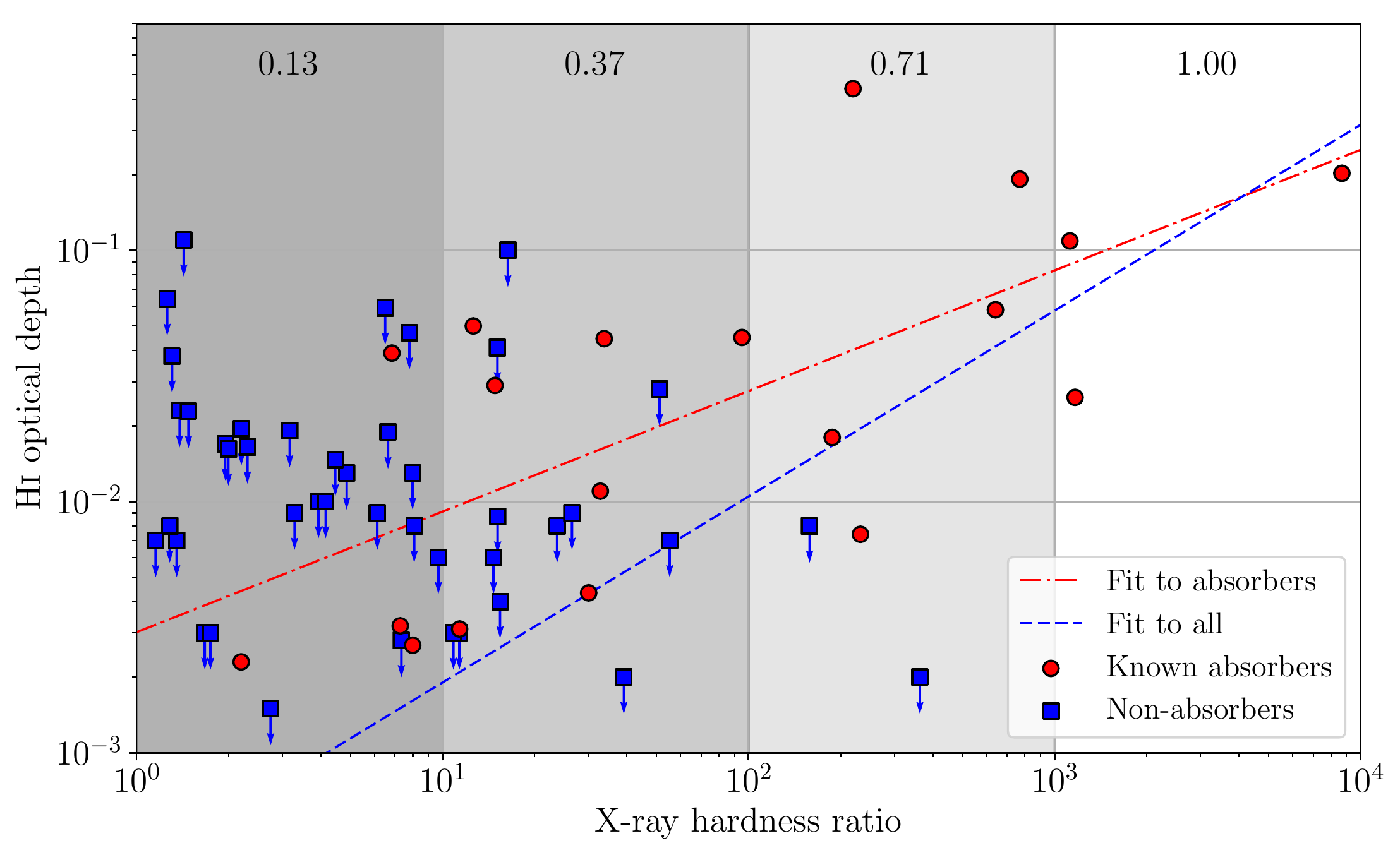}
\caption{Optical depth $\tau$ (or 3$\sigma$ upper limit on $\tau$ for non-absorbers) versus X-ray hardness ratio for 67 galaxies in our sample. As X-ray ratio increases, the fractional proportion of H{\sc i} absorbers (red circles) also increases (shown as text labels for each grey-shaded order of magnitude bin in X-ray ratio) relative to the number of non-absorbers (blue triangles). We also plot the trends for the detections (red dashed line) and the whole sample including upper limits as determined by survival analysis (blue solid line), confirming a trend at 4.71$\sigma$ in optical depth versus X-ray hardness ratio.}
\label{fig:tauratio}
\end{figure}

\begin{figure}
\includegraphics[angle=0,width=0.46\textwidth]{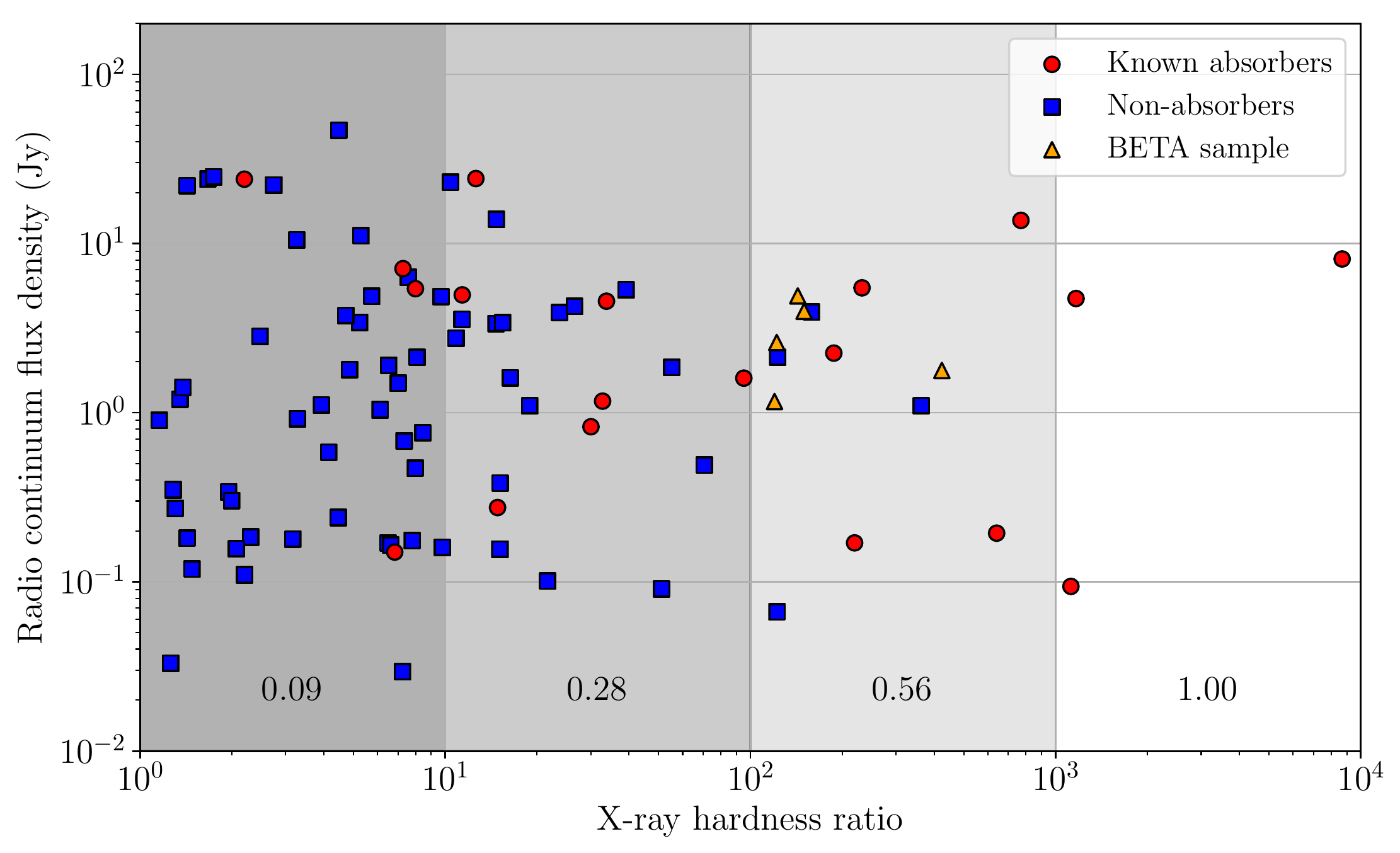}
\caption{Radio continuum flux versus X-ray hardness ratio for both the sample of \atot galaxies (shown as red circles for H{\sc i} absorbers and blue squares for non-absorbers) and for our BETA sample (orange triangles). We chose our BETA sample to occupy the high end of X-ray ratio parameter space (where the fraction of absorbers is high) and to be brighter than 1\,Jy in radio emission.}
\label{fig:fluxratio}
\end{figure}

We also consider the relationship between X-ray ratio and the chance of a detection of H{\sc i} absorption. Figure \ref{fig:tauratio} shows 21\,cm optical depth $\tau$ (either the peak $\tau$ of an absorber or the upper limit on a non-absorber) versus X-ray hardness ratio for a reduced sample of 67 galaxies (those for which we have measured 21\,cm optical depths or upper limits). As the X-ray hardness ratio increases, the fraction of galaxies that have H{\sc i} absorption also increases from 0.1 to 1 (shown as text labels in the figure, although it is relatively small number statistics at the high end). To characterise the trends in each population, we carried out survival analysis \citep{Feigelson:1985by,Isobe:1986iu} on the reduced sample of 67 galaxies. A generalised non-parametric Kendall-tau test (where tau here is distinct from optical depth $\tau$) on the detected 21\,cm points gives a probability of $P(\tau) = 2.30\times10^{-3}$ of the distribution occurring by chance, which is significant at $S(\tau) = 3.04\sigma$ assuming Gaussian statistics. We can include the upper limits to the optical depths, from the non-detections, via the {\em Astronomy SURVival Analysis} ({\sc asurv}) package \citep{Isobe:1986iu,Lavalley:1992wf}. We note that these upper limits represent the 3$\sigma$ confidence interval. These are added to the \adet detected sight-lines as censored data points, giving $P(\tau) = 2.38\times10^{-6}$ for the whole sample, which is significant at $S(\tau) = 4.71\sigma$. Thus, the data indicates a strong correlation between the optical depth and the X-ray hardness ratio. We plot these data, as well as the fits determined by the survival analysis, in Figure \ref{fig:tauratio}. We show a similar plot, but for radio continuum flux instead of optical depth, in Figure \ref{fig:fluxratio}. This figure includes all of our \atot galaxy sample, as well as our pilot BETA sample (shown as orange triangles). The fraction of H{\sc i} absorbers in each order-of-magnitude bin is again shown but for the complete sample (excluding our BETA sample), with the same increase in fraction as X-ray ratio increases.

Overall, based on this sample of \adet H{\sc i} absorbers and \anon non-absorbers, we find that the X-ray hardness ratio $\frac{f_5}{f_1}$ derived from {\it XMM-Newton} presents a useful diagnostic for increasing the chance of an H{\sc i} absorption detection. We also find in this limited sample evidence for a positive correlation between optical depth and X-ray ratio at a significance of 3.04$\sigma$ for the detections and 4.71$\sigma$ including upper limits, affirming the notion that these different probes of hydrogen near an AGN are tracing similar, if not the same, spatial regions of gas. With a larger sample of H{\sc i} absorbers and non-absorbers, such as we expect to obtain with the upcoming FLASH search towards 150,000 background galaxies or via other large-scale H{\sc i} absorption line surveys, we can confirm this trend with even higher statistical significance. 

\subsection{Manual reprocessing of X-ray data}
Our analysis above was based on using the flux values as reported in the 3XMM-DR4 catalogue, which are measured by running a standard pipeline on all serendipitously-detected sources in a given {\it XMM-Newton} observed field. The subtleties of reducing data for individual but varied sources and potential impact on the resulting flux values means that we must gauge the reliability of the pipeline-produced catalogue versus the manually-intensive but more accurate procedure of reducing each observation by hand. As such, we tested the accuracy of the flux values of the 3XMM-DR4 catalogue by manually reprocessing {\it XMM-Newton} data from a selection of sources from Table 1. All \adet H{\sc i} absorbers were selected, while a random selection of 11 non-H{\sc i} absorbers across a range in flux were used. Fewer non-H{\sc i} absorbers were investigated as they were found to be more consistent with the catalogue data.

{\it XMM-Newton} data were downloaded from NASA's High Energy Astrophysics Science Archive Research Centre (HEASARC) archive. For each source, we selected the single longest publicly-available European Photon Imaging Camera (EPIC) observation and re-processed the data using standard tasks from the Science Analysis System ({\small SAS}) version 14.0.0 software tool. Firstly, we excluded high particle background exposure time intervals. We used the standard flagging criterion \verb|FLAG=0|, along with \verb|#XMMEA_EM| and \verb|#XMMEA_EP| for the MOS and pn detectors respectively. Patterns 0-12 were used for MOS and 0--4 for pn. Background-subtracted spectra were extracted using source regions of 35$''$ radius (for crowded fields, smaller regions were used) and local background regions at least three times larger and of similar distance from the readout nodes. We used the {\small SAS} task {\it epatplot} to identify any sources significantly affected by pileup. Three sources that were badly affected by pileup were excluded from the comparison sample as they would have been subject to increased uncertainties. Individual spectra for MOS1, MOS2 and pn were extracted and then combined using {\it epicspeccombine} to create a weighted-average EPIC spectrum. This spectrum was then grouped to at least 20 counts per bin in order to use $\chi^2$ statistics. We used XSPEC version 12.8.2 \citep{1996ASPC..101...17A} for all spectral analysis. 

\begin{figure}
    \centering
    \includegraphics[angle=0,width=0.46\textwidth]{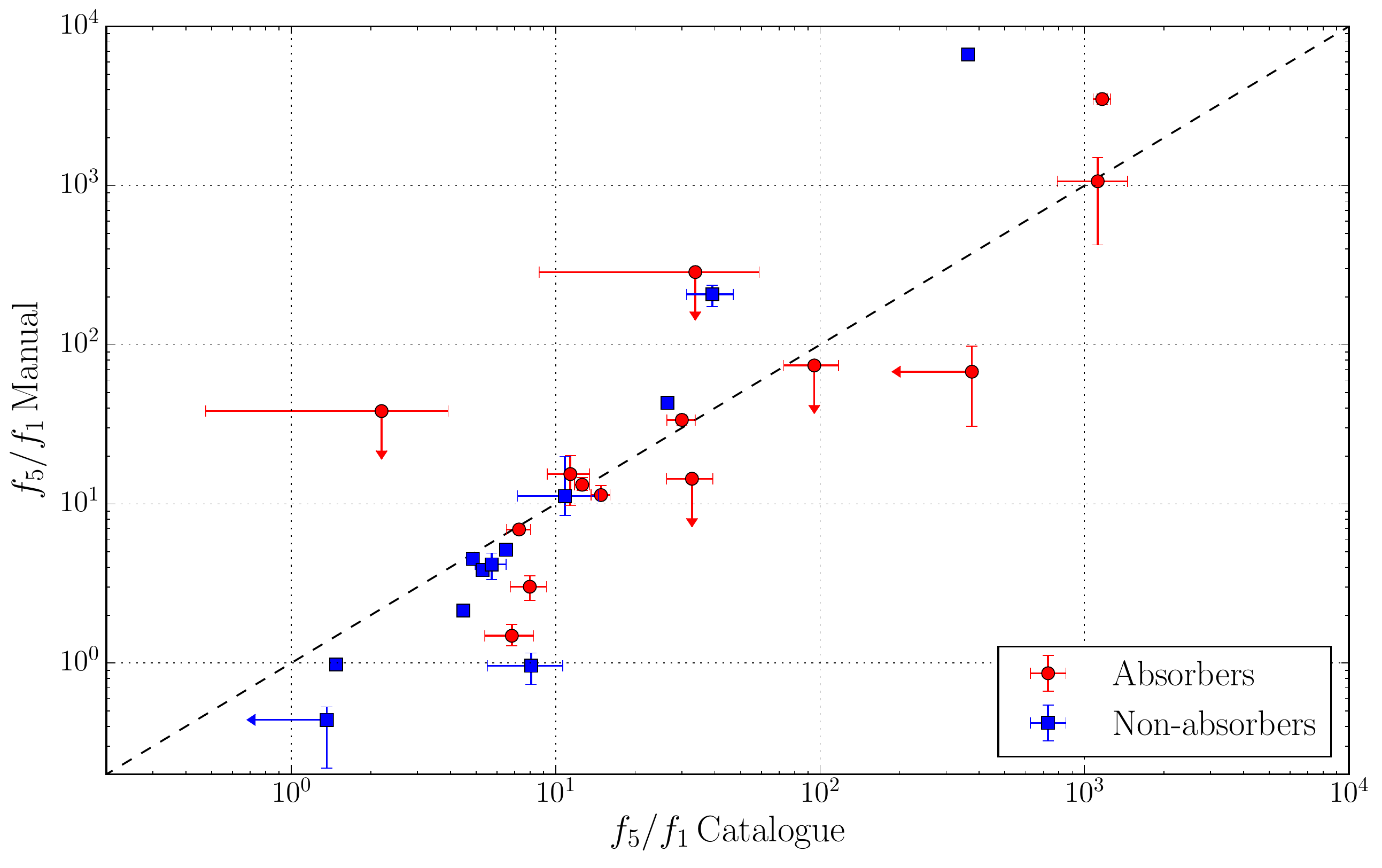}
    \caption{Comparison of flux ratios ($f_5/f_1$) between catalogue-derived values and manually extracted values. Red circles represent H{\sc i} absorbing sources while blue squares represent non-absorbing sources. Arrows indicate upper limits. The black dotted line represents catalogue-listed fluxes equal to fitted fluxes. Ignoring upper limits, we find a Spearman rank of $r = 0.86 \pm 0.04$.}
    \label{comparison_fig}
    \vspace{0.3cm}
\end{figure}

Each source was initially fitted with a {\it phabs $\times$ zphabs $\times$ pexriv} \citep{1995MNRAS.273..837M} model, describing non-intrinsic photoelectric absorption, photoelectric absorption at the redshift of the galaxy, and a power-law reflection component. Here, the {\it phabs} component is fixed to the line-of-sight absorption contribution while the {\it zphabs} component is left free (with redshift fixed to its respective host redshift). For most of the non-absorbers, this model is sufficient to replicate the continuum. For some of the absorbers, additional components, such as {\it mekal} (hot diffuse gas) and {\it reflect} (reflection from neutral material) components, were required to obtain good fits. Using the {\it cflux} component, the fluxes in bands 1 (0.2--0.5\,keV) and 5 (4.0--12.0\,keV) were determined. Finally a ratio of these two fluxes was calculated and compared with the catalogue equivalent for each of our selected sources (Figure \ref{comparison_fig}). Removing upper limits, we find a strong correlation between the manually extracted and catalogue flux values (Spearman rank $r = 0.86 \pm 0.04$ using a Monte Carlo perturbation error analysis). 

Five significant outliers were ignored for this analysis. One of these, H{\sc i} absorber J125711$-$172434, has a hardness ratio from the catalogue of $\sim$1 which upon inspection of the data is due to its location at the centre of a dense cluster (Abell\,1644). The galaxy is embedded in very hot `sloshing' cluster gas at a luminosity of $\sim$10$^{42}$\,erg\,s$^{-1}$ \citep{Johnson:2010ix}, likely to be dwarfing emission from the AGN and thus responsible for its observed flat spectrum. Due to the complicated nature of this particular system and difficulty in extracting meaningful information from the spectral data, we excluded it from this analysis. The other four outliers were all highly absorbed, and likely Compton-thick, sources ($N_{\rm H} > 10^{24}$\,cm$^{-2}$). In the catalogue, the flux in the soft band for those sources is extremely low ($f_1 \ll f_5$) but non-negligible. However, our individual modelling of the higher-energy continuum suggests that the direct flux from the AGN should be completely absorbed (below the detection limit) in the softest band. Thus, the residual emission detected in that band must have a different origin: either from the host galaxy (at much lower $N_{\rm H}$) or from the AGN, but scattered into our line of sight above the obscuring material. In any case, the catalogue value of the hardness ratio $\frac{f_5}{f_1}$ will be an underestimate of the true hardness ratio for the direct AGN emission component. Although this only occurs in a small number of sources, and seems to be restricted to H{\sc i} absorbing sources, we still advise caution when using the X-ray flux ratios to estimate H{\sc i} absorption, specifically for strong absorbers. Future methods that we adopt for selecting X-ray absorbed sources will be developed on the basis of the caveats revealed by our analysis here, however we conclude that the simple measure of X-ray hardness ratio $\frac{f_5}{f_1}$ is a useful method to first order to select for likely H{\sc i} absorbers.

\section{ASKAP-BETA test observations of X-ray absorbers}
In order to test correlation between soft X-ray absorption and H{\sc i} absorption, we compiled a list of candidates to be potentially observed with BETA. We began with a broader set of criteria defining a set of southern-sky sources, which we then narrowed to meet further redshift requirements specific to BETA. Our initial broad criteria were:

\begin{table*}
\centering
\caption{Properties of our complete X-ray sample. A total of 11 southern-sky sources were identified on the basis of radio brightness ($>$ 1\,Jy), X-ray hardness ($>$ 100), declination $\delta < 0^\circ$ and Galactic latitude $|b| \ge 5^\circ$. The radio flux densities were obtained from either NVSS at 1.4\,GHz \citep{Condon:1998kn} or SUMSS \citep{Mauch:2003ed} at 843\,MHz. Of these, 3 galaxies are known H{\sc i} absorbers (5 including PKS\,1740$-$517 and PKS\,1657$-$298). For observations with BETA, redshifts were either unknown or selected to be in the range 0.4 $<$ $z$ $<$ 1.0. The sample observed with BETA, including PKS\,1740$-$517, occupies the bottom half of this table. The last column indicates the presence or absence of H{\sc i} absorption; we note that we can only rule out H{\sc i} absorption for our BETA sample within our searched redshift range.}\label{tb:sample}
\begin{tabular}{lcccccccccc}
\hline
 Source & XMM name &RA, Dec & S$_{1.4}$ &  S$_{0.8}$&  X-ray ratio & $z$ & $R$-band & H{\sc i} abs.\\ 
&&J2000&Jy&Jy&&&mag \\
\hline
PKS\,0625$-$545		& J062648.7$-$543208	&	06:26:49.58, $-$54:32:34.7$^a$		& -			&	1.15	&	487		&	0.052$^a$		&	14.5$^p$ & ? \\
PKS\,0859$-$25		& J090147.2$-$255516	&	09:01:47.50, $-$25:55:18.7$^b$		& 6.25		&	-		&	481		&	0.305$^i$		&	18.9$^p$ & ? \\
NGC\,4945			& J130527.4$-$492804	&	13:05:27.48, $-$49:28:05.6$^c$		& -			&	5.55	&	3764	&	0.002$^j$		&	9.5$^q$ & Y \\
PKS\,1549$-$79  	& J155658.9$-$791403	&	15:56:58.87, $-$79:14:04.3$^d$		& -			&	6.19	&	1167	&	0.152$^k$		&	17.9$^p$ & Y \\
PKS\,1814$-$63  	& J181935.0$-$634547	&	18:19:35.00, $-$63:45:48.2$^d$		& -			&	19.89	&	770		&	0.063$^m$		&	15.6$^p$ & Y\\
\hline
PKS\,1547$-$79		& J155521.7$-$794036	&	15:55:21.65, $-$79:40:36.3$^e$		&	-		&	6.01 	& 	143		&	0.483$^n$	 	& 19.1$^p$ & N\\
PKS\,1657$-$298	& J170109.8$-$295439	&	17:01:09.71, $-$29:54:41.8$^b$		&	1.78	& -			&  	424		&	0.420$^r$ 				&- & Y\\
MRC\,1722$-$644	& J172657.7$-$642751	&	17:26:57.85, $-$64:27:53.4$^f$		&	-		&	4.14	&	150		&	- 				&- & N\\
PKS\,1740$-$517  	& J174425.3$-$514444	&	17:44:25.45, $-$51:44:43.8$^d$		& -			&	8.15	&	8689	&	0.441$^l$		&	20.8$^n$ & Y\\
PKS\,2008$-$068	& J201114.2$-$064403	&	20:11:14.23, $-$06:44:03.4$^g$		&	2.60	&	- 		&	122		&	0.547$^o$ 		& *21.2$^o$& N\\
PKS\,2154$-$11		& J215644.1$-$112747	&	21:56:44.10, $-$11:27:49.5$^h$		&	1.16	&	-		& 	120		&	- 				&- & N\\
\hline
\end{tabular}\\
$^a$\citet{Cava:2009eo} $^b$\citet{Douglas:1996jm} $^c$\citet{Greenhill:1997ci} $^d$\citet{Johnston:1995js} $^e$\citet{diSeregoAlighieri:1994tr} $^f$\citet{Murphy:2010ej} $^g$\citet{Mahony:2011ea} $^h$\citet{Condon:1998kn}  $^i$\citet{Burgess:2006df} $^j$\citet{Koribalski:2004cv} $^k$\citet{Holt:2008kz} $^l$\citet{2015MNRAS.453.1249A} $^m$\citet{RamosAlmeida:2011ga} $^n$\citet{Tadhunter:1993kt} $^o$\citet{Snellen:2002hh} $^p$\citet{Hambly:2001dq} $^q$\citet{Doyle:2005jt} $^r$this work\\
*Observed with $r$-band filter at 6814\,\AA~by \citet{Snellen:2002hh} 
\end{table*}

\begin{enumerate}[~~1)]
\item Radio flux $S$ $>$ 1\,Jy (in SUMSS or NVSS)
\item X-ray hardness ratio $\frac{f_5}{f_1}$ $>$ 100 in 3XMM-DR4
\item Declination $\delta$ $<$ 0$^\circ$
\item Galactic latitude $|b| \ge 5^\circ$
\end{enumerate}

\begin{figure}
\centering
\includegraphics[angle=0,width=0.45\textwidth]{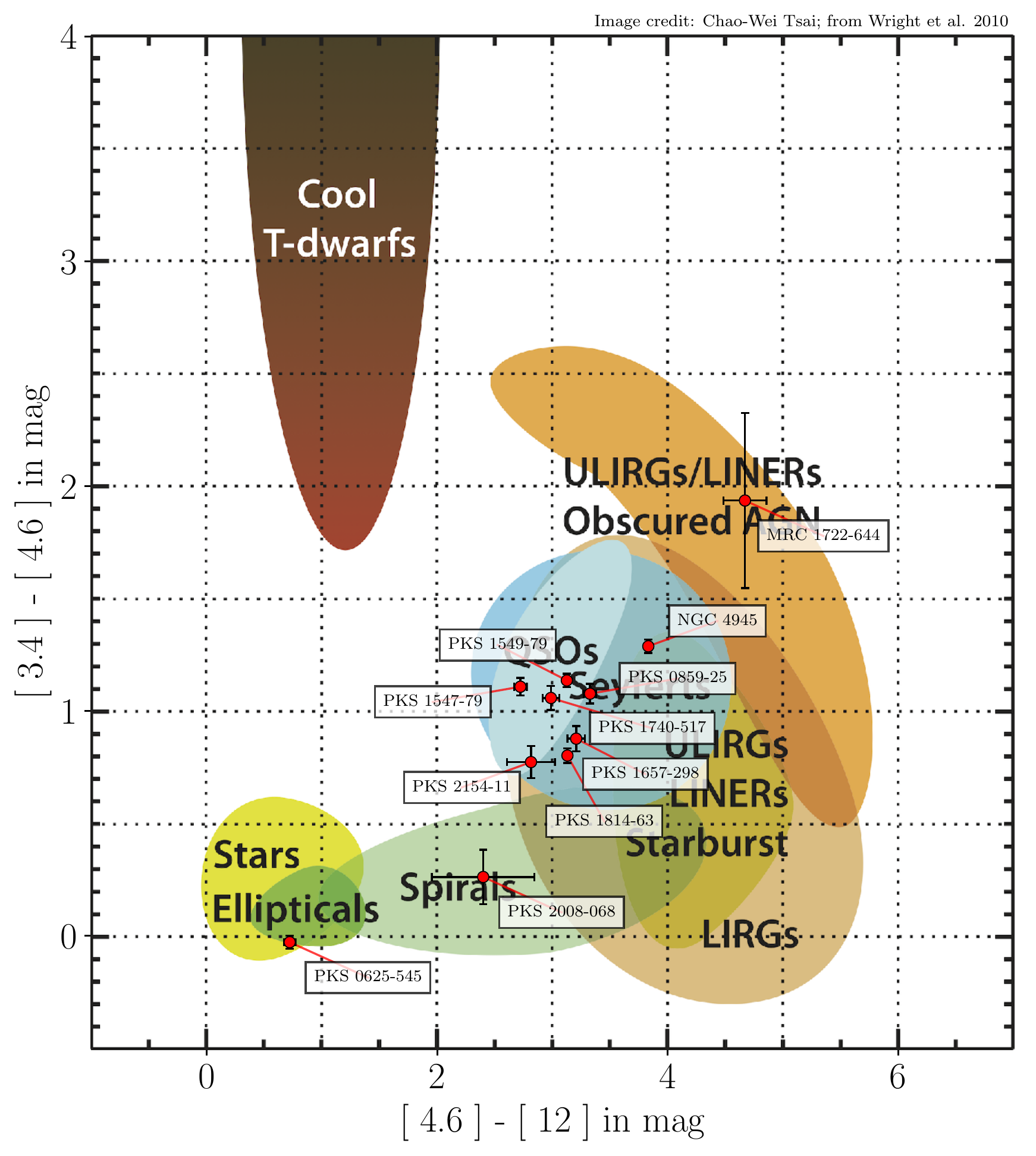}
\caption{The 11 radio sources in our total sample (including the 5 sources presented in this paper) plotted on the \citet{Wright:2010in} WISE colour-colour diagram to give an insight into the properties of their likely host galaxies. Our sources occupy a range in possible galaxy type, including elliptical, spiral, Seyfert, QSO and ULIRG.}
\label{fig:wiseplot}
\end{figure}

The adoption of these first criteria was primarily driven by the sensitivity limitations of BETA in its 6-antenna configuration. This made it necessary to target the brightest radio sources ($S >$ 1\,Jy) in order to reach a reasonable optical depth sensitivity within an initial 3\,hr observation. Because there were no candidates that were both brighter than 1\,Jy in radio and with X-ray ratios $>$ 1000 that were not already known H{\sc i} absorption detections, we lowered the X-ray ratio threshold to investigate whether ratios $>$ 100 still predicted H{\sc i} absorption in the same way as the most-absorbed X-ray spectra, with the expectation that we were less likely to detect H{\sc i} absorption in this parameter space than those sources with higher ratios. We note that applying these criteria results in a total of 11 candidate sources (7 from SUMSS, 4 from NVSS). However, of these, four sources are known H{\sc i} absorbers (NGC\,4945, PKS\,1549$-$79, PKS\,1740$-$517, PKS\,1814$-$63) while an additional two sources are beyond the redshift range we targeted with BETA (PKS\,0625$-$545 at $z$ = 0.05 and PKS\,0859$-$25 at $z$ = 0.3). Nonetheless, this full sample provides a useful parent set based on clear criteria from which we draw our BETA subset. An overview of the properties of the entire sample is given in Table \ref{tb:sample} along with the respective references. The top half of Table \ref{tb:sample} contains sources with $z < 0.4$, of which the three galaxies with the highest X-ray hardness ratios already have known detections of H{\sc i} absorption while the other two have yet to be searched. The bottom half of Table \ref{tb:sample} contains our BETA sample detailed in this paper, as well as PKS\,1740$-$517 described in \citet{2015MNRAS.453.1249A}; in the same way, the two sources with the highest X-ray hardness ratios also feature H{\sc i} absorption, which is consistent with our findings in Section \ref{propcompare}. Each source is detected in the mid-infrared, and so we include their positions overplotted on the \citet{Wright:2010in} WISE colour-colour plot in Figure \ref{fig:wiseplot} to give insight into the likely properties of their host galaxies. From this, we see that this sample spans a range in galaxy types based on WISE colours, although mostly occupying the region near the QSO class. 

The final list of candidates observed with ASKAP-BETA (the bottom half of Table \ref{tb:sample}) were required to meet the following additional criterion:

\begin{enumerate}[~~1)]\addtocounter{enumi}{4}
\item (if known) redshift $z$ in the range $0.4 \le z \le 1.0$
\end{enumerate}

For our observed sample we focused on the redshift range 0.4 $\le z \le$ 1.0 as this corresponds to Band 1 of BETA (711.5--1015.5\,MHz), which was the band primarily used during commissioning, allows access to the most redshift parameter space at once, is the least affected by RFI and is the most unexplored in terms of H{\sc i} absorption. This leaves our BETA X-ray pilot sample at a total of five sources (six including PKS\,1740$-$517), described in more detail in the next section. We do not discuss the properties of PKS\,1740$-$517 here, and instead refer the reader to \citet{2015MNRAS.453.1249A} for details about this source. 

\subsection{Properties of our observed sample}\label{betaprop}
Here we summarise what has been previously reported about each source in our pilot sample. We show the radio spectral energy distribution for each source in Figure \ref{fig:seds} to give an insight into their radio continuum properties, with detail given in Section \ref{modelling}. We searched for potential optical counterparts in the NASA/IPAC Extragalactic Database (NED) and SuperCOSMOS blue/red surveys (within 5$''$). We also checked to see if the sources have been observed at high angular resolution in radio, using very long baseline interferometry (VLBI). 

\subsubsection{PKS\,1547$-$79}
A source known to be linearly polarised at a level $>$ 5\% across the range from 10-30\,cm wavelengths \citep{Gardner:1966wg}, it was also included in the \citet{Kuehr:1981wo} sample based on its 5\,GHz flux density being $>$ 1\,Jy. It was imaged at 2$''$ resolution by \citet{Duncan:1992ut} revealing a double-lobed structure and central radio source. It is part of the 2\,Jy sample of \citet{Tadhunter:1993kt} who confirmed an optical redshift of $z$ = 0.483, and has been identified as a broad line radio galaxy (BLRG) likely interacting with nearby companions. PKS\,1547$-$79 was also part of an {\it XMM-Newton} study conducted by \citet{Mingo:2014bx}, who found a high intrinsic $N_{\rm H}$ from X-ray absorption of $\sim$9.9 $\times$ 10$^{23}$\,cm$^{-2}$ potentially indicating a Compton-thick medium surrounding the AGN. 

\subsubsection{PKS\,1657$-$298}
This is a largely unknown and unstudied radio galaxy, identified as a compact double at VLBI-resolution. Due to its low Galactic latitude ($+7^\circ$), there has been no optical counterpart identified and no previously known redshift for the galaxy. During a study of {\it ROSAT} X-ray sources in globular clusters, \citet{2001A&A...368..137V} identified the X-ray counterpart for this galaxy within the field of view to a positional accuracy of 3$''$. \citet{Petrov:2012if} provide the most accurate position for this source based on VLBI data at 22\,GHz.

\begin{figure*}
\includegraphics[angle=0,width=0.46\textwidth]{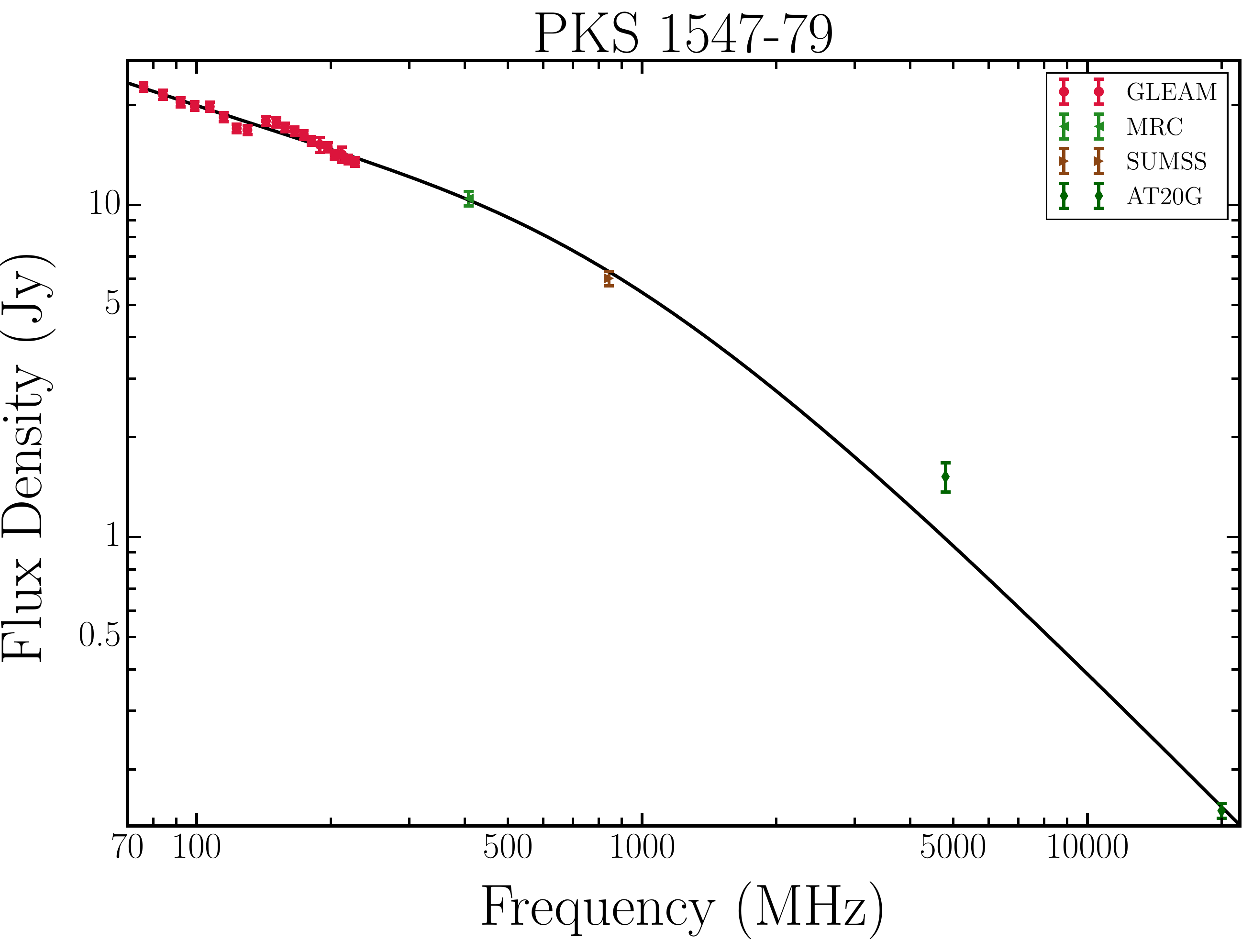}
\includegraphics[angle=0,width=0.46\textwidth]{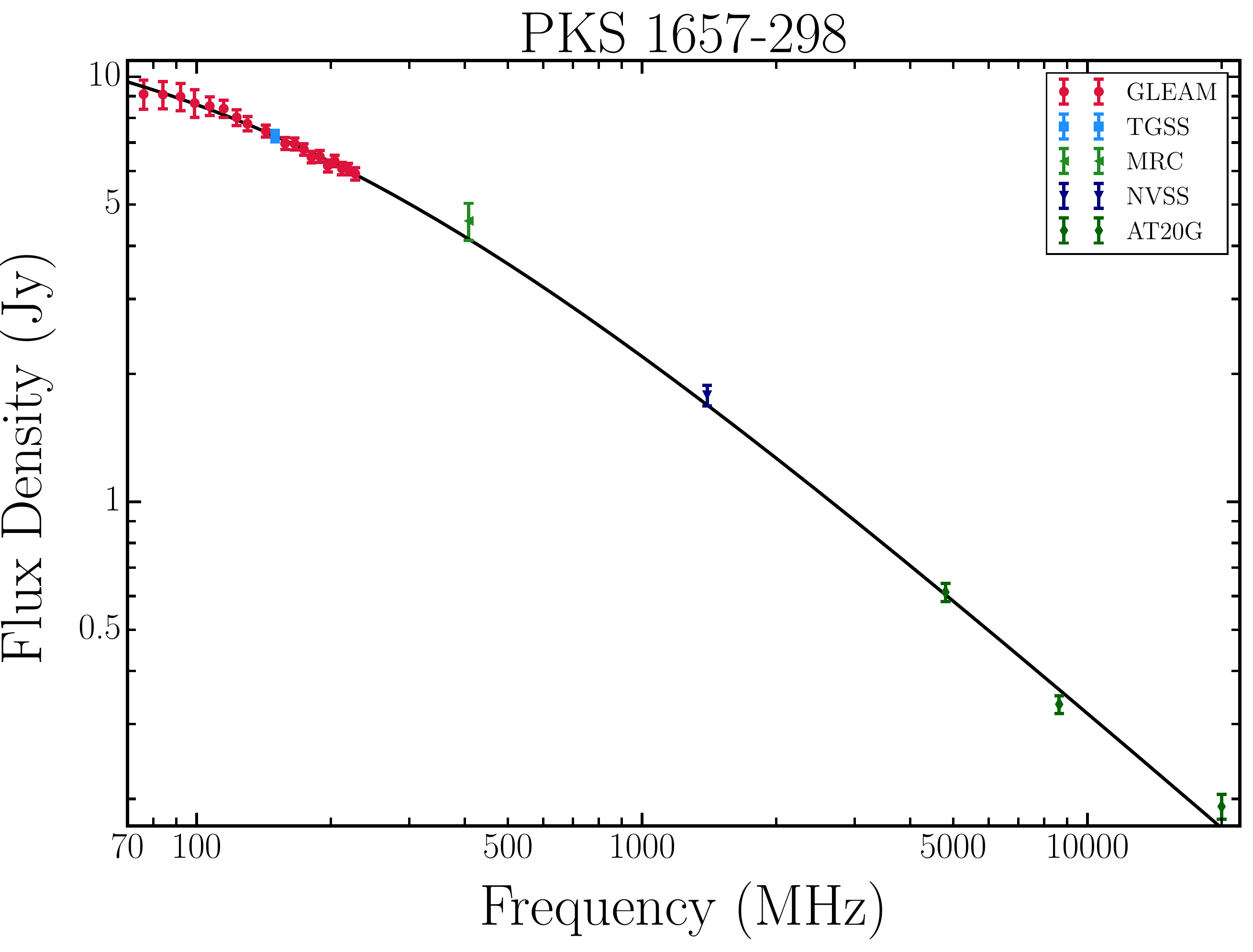}
\includegraphics[angle=0,width=0.46\textwidth]{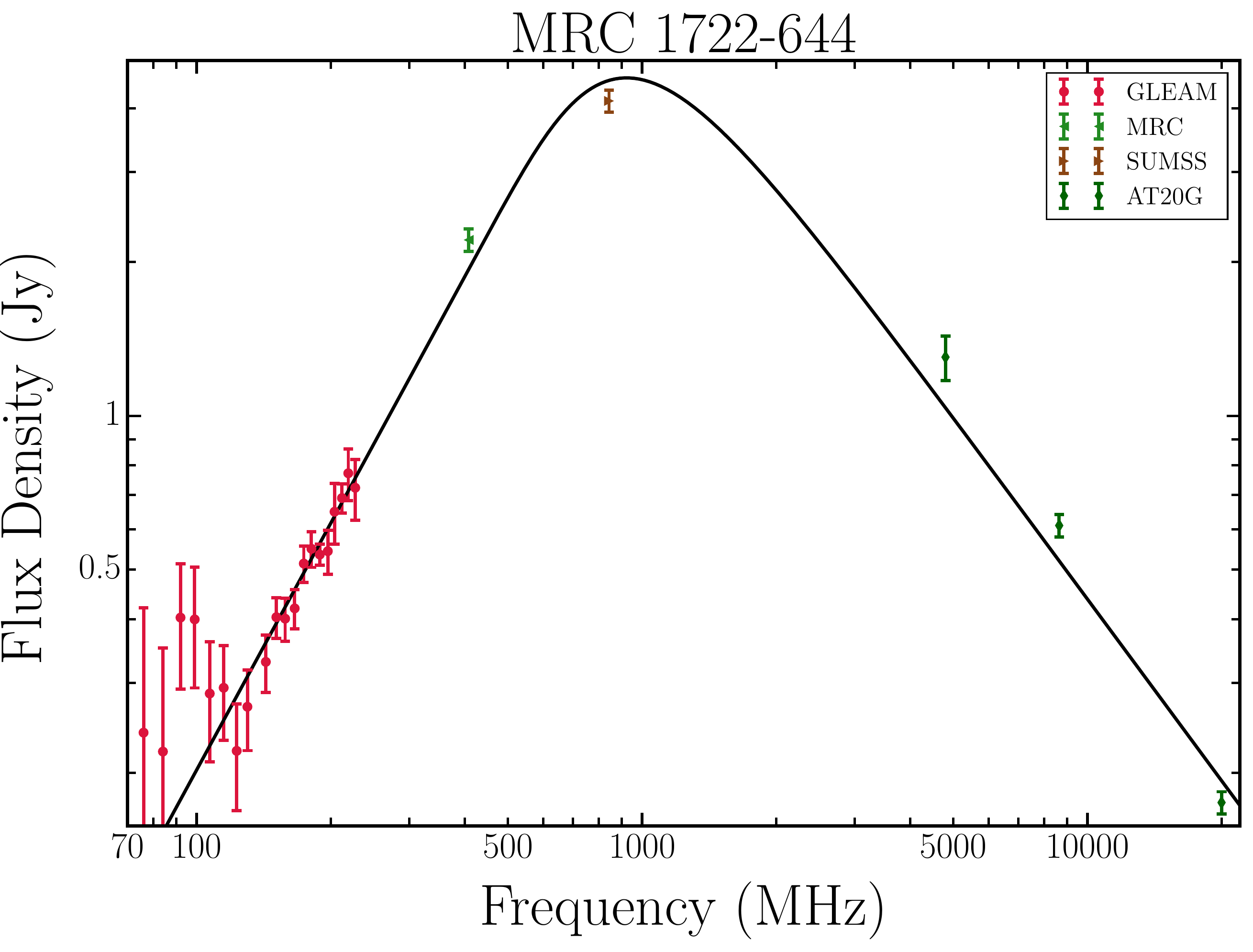}
\includegraphics[angle=0,width=0.46\textwidth]{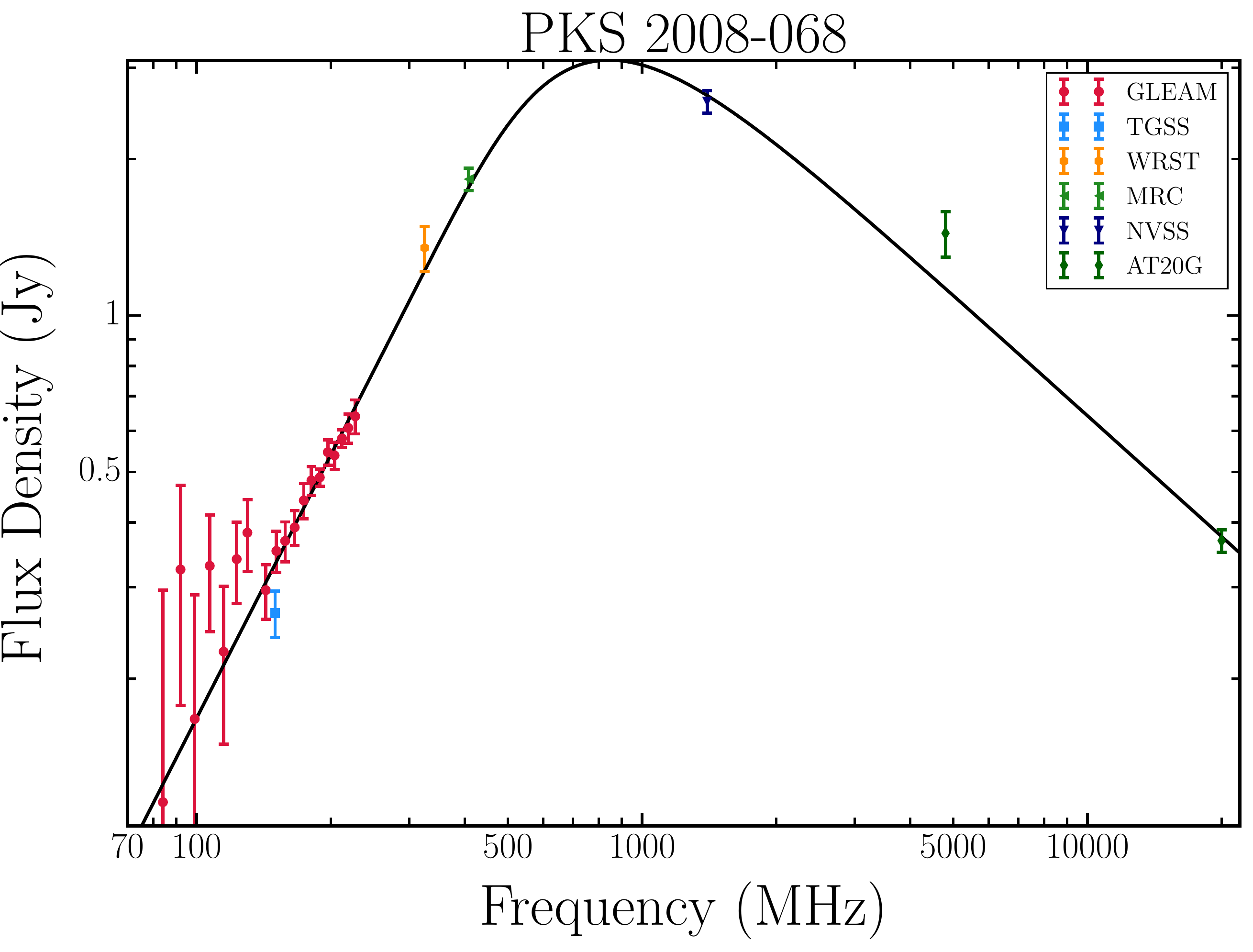}
\includegraphics[angle=0,width=0.46\textwidth]{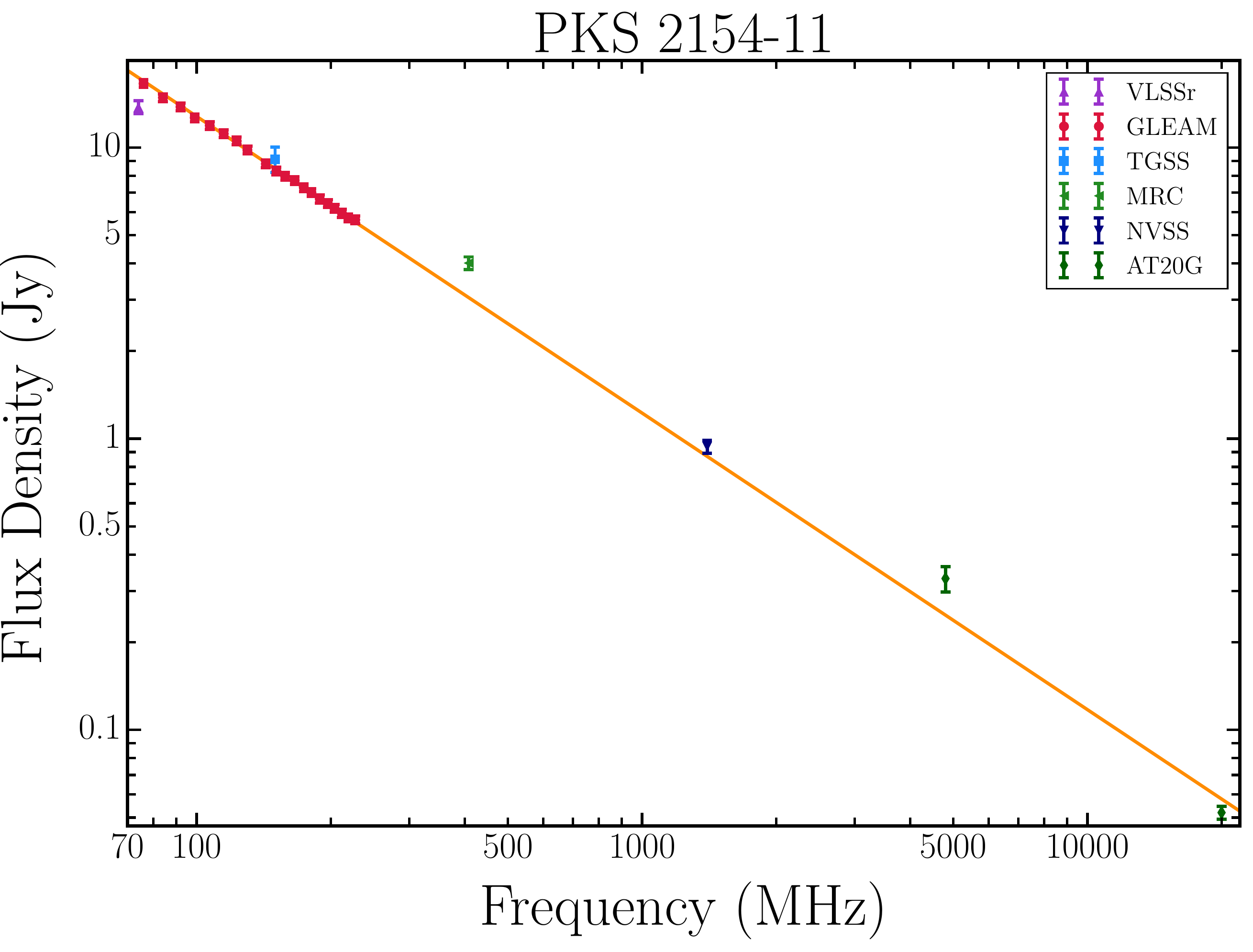}
\caption{Radio spectral energy distributions for the BETA sample. The black curve represents the fit of Equation \ref{eqn:gen_curved} to all of the available data. PKS\,2154$-$11 shows no evidence of curvature in its spectrum, and thus is fit with a standard power-law fit, as shown by the orange line. The surveys appearing here include (ordered from lowest to highest frequency): the Very Large Array Low-Frequency Sky Survey redux at 75\,MHz \citep[VLSSr,][]{Lane:2014cr}, the GaLactic and Extragalactic All-sky MWA survey at 72--231\,MHz \citep[GLEAM,][]{Wayth:2015kp,HurleyWalker:2017kq}, the TIFR GMRT Sky Survey ADR1 at 150\,MHz \citep[TGSS-ADR1,][]{Intema:2017bd}, a sample of GPS sources observed with the Westerbork Synthesis Radio Telescope at 327\,MHz \citep[WSRT,][]{Stanghellini:1998gb}, the Molonglo Reference Catalogue at 408\,MHz \citep[MRC,][]{Large:1981vm}, the Sydney University Molonglo Sky Survey at 843\,MHz \citep[SUMSS,][]{Mauch:2003ed}, the NRAO VLA Sky Survey at 1.4\,GHz \citep[NVSS,][]{Condon:1998kn}, and the Australia Telescope 20\,GHz survey \citep[AT20G,][]{Murphy:2010ej}. }
\label{fig:seds}
\end{figure*}

\subsubsection{MRC\,1722$-$644}
First identified in MRC at 408\,MHz as 1722$-$644, this is a radio source of mostly unknown multi-wavelength properties. It is a compact double at VLBI-resolution \citep{Edwards:2016jw}, and has no known redshift or optical counterpart. This source is part of the GPS class of radio sources, due to its spectral peak at ~1\,GHz, and features a very narrow radio spectrum \citep{Edwards:2004ix}. 

\subsubsection{PKS\,2008$-$068}
One of the more well-studied sources in our sample, this galaxy falls into the category of GPS sources with its peak at 1.6\,GHz. Similar to our other sources, it is compact on VLBI-scales. It is known to be a weakly-polarised source \citep{Berge:1972hr}, at a level of 1\% at 10\,GHz \citep{Inoue:1995us}. It has a confirmed optical redshift $z$ = 0.547 \citep{Snellen:2002hh}.

\subsubsection{PKS\,2154$-$11}
A mostly obscure radio source with little known about the properties of the host galaxy. It has been detected across the radio spectrum from 80\,MHz to 20\,GHz, and was reported as extended at 20\,GHz \citep{Murphy:2010ej}. Its WISE colours place it in the QSO class of Figure \ref{fig:wiseplot}, but there is no known optical counterpart and no reported redshift.

\begin{table*}
	\small
	\caption{\label{table:fit_params} The best fitting statistics of the generic curved model from Equation \ref{eqn:gen_curved} for PKS\,1547$-$79, PKS\,1657$-$298, MRC\,1722$-$644, and PKS\,2008$-$068. Only the parameters from the non-thermal power-law fit are presented for PKS\,2154$-$11 since it is best fit by that model. Note that the curvature evident in the spectrum of PKS\,1547$-$79 is most likely the product of a spectral break occurring near 1.4\,GHz, or the spectrum becoming dominated by core emission, rather than an absorption process.}
	\begin{center}
		\begin{tabular}{cccccc}
		\hline
		\hline
Name & $S_{\mathrm{p}}$\,(Jy)  & $\nu_{\mathrm{p}}$\,(MHz) & $\alpha_{\mathrm{thick}}$ &  $\alpha_{\mathrm{thin}}$ & $a$\,(Jy) \\
		\hline			
PKS\,1547$-$79 & $4.0 \pm 0.5$ & $1390 \pm 150$ & $-0.4 \pm 0.1$ & $-1.4 \pm 0.2$ & ... \\
PKS\,1657$-$298 & $9.1 \pm 0.3$ & $100 \pm 30$ & $0.1 \pm 0.2$ & $-0.9 \pm 0.2$ & ... \\
MRC\,1722$-$644 & $4.6 \pm 0.1$ & $930 \pm 40$ & $1.6 \pm 0.1$ & $-1.2 \pm 0.1$ & ...\\
PKS\,2008$-$068 & $3.0 \pm 0.2$ & $740 \pm 10$ & $1.7 \pm 0.1$ & $-0.8 \pm 0.1$ & ... \\
PKS\,2154$-$11 & ... & ... & ... & $-1.0 \pm 0.1$ & $1.4 \pm 0.2$\\
		\hline\end{tabular}                           
\end{center}                                                                               
\end{table*}

\subsection{Spectral modelling at low radio frequencies}\label{modelling}
The sources in which we searched for H\textsc{i} absorption with BETA were also observed as part of the low radio frequency all-sky survey conducted by the Murchison Widefield Array \citep[MWA,][]{Tingay:2013ef}. The GaLactic and Extragalactic All-sky MWA survey \citep[GLEAM,][]{Wayth:2015kp,HurleyWalker:2017kq} observed the entire sky south of $+30^\circ$ declination at twenty different frequencies between 72 and 231\,MHz. The extragalactic catalogue formed from the GLEAM survey consists of 307,455 sources, with positions accurate to $\sim$\,30$''$. This low radio frequency data allows us to constrain the spectral shape of the sources and, in cases where free-free absorption is responsible for a spectral turnover, infer the density of the ionised medium in a source \citep[e.g.][]{Callingham:2015hta}.

We fit the following generic curved model to characterise the radio spectrum of the five sources searched for H\textsc{i} absorption with BETA:

\begin{equation}\label{eqn:gen_curved}
 S_{\nu} = \dfrac{S_{\mathrm{p}}}{(1-e^{-1})}\left[1 - e^{-\left(\nu / \nu_{\mathrm{p}}\right)^{\alpha_{\mathrm{thin}} - \alpha_{\mathrm{thick}}}}\right]\left(\dfrac{\nu}{\nu_{\mathrm{p}}}\right)^{\alpha_{\mathrm{thick}}},
\end{equation} 

\noindent where $\alpha_{\mathrm{thick}}$ and $\alpha_{\mathrm{thin}}$ are the spectral indices in the optically thick and optically thin regimes of the spectrum, respectively. $S_{\mathrm{p}}$ is the flux density at the peak frequency $\nu_{\mathrm{p}}$ \citep{Snellen:1998jk}. The standard non-thermal synchrotron power-law of the form

\begin{equation}\label{eqn:powlaw}
 S_{\nu} = a\nu^{\alpha},
\end{equation} 

\noindent where $a$ characterises the amplitude of the synchrotron spectrum, was also fit to the spectrum if the source did not display significant spectral curvature. The procedure implemented here to fit the spectra is identical to that outlined in \citet{Callingham:2016hv}. In summary, a Markov Chain Monte Carlo algorithm was used to sample the posterior probability density functions of the various model parameters such that the applied Gaussian likelihood function was maximised under physically sensible uniform priors. The simplex algorithm introduced by \citet{Nelder:1965in} was used to direct the walkers to the maximum of the likelihood function.

PKS\,1547$-$79, PKS\,1657$-$298, MRC~1722$-$644, and PKS\,2008$-$068 all display significant curvature in their spectra between 72\,MHz and 20\,GHz, and are best fit by Equation \ref{eqn:gen_curved}. Only PKS\,2154$-$11 is best fit by the simple power-law of Equation \ref{eqn:powlaw}. The best fitting statistics for the five sources are presented in Table \ref{table:fit_params}, and the data are plotted in Figure \ref{fig:seds}. Note that the spectral peak for PKS\,1657$-$298 is only just detected within the GLEAM band, which means the optically thick spectral index for this source has a large uncertainty since the spectrum is not completely sampled below the turnover.

\subsection{BETA observations and data reduction}
\begin{figure*}
\centering
\includegraphics[angle=0,width=0.8\textwidth]{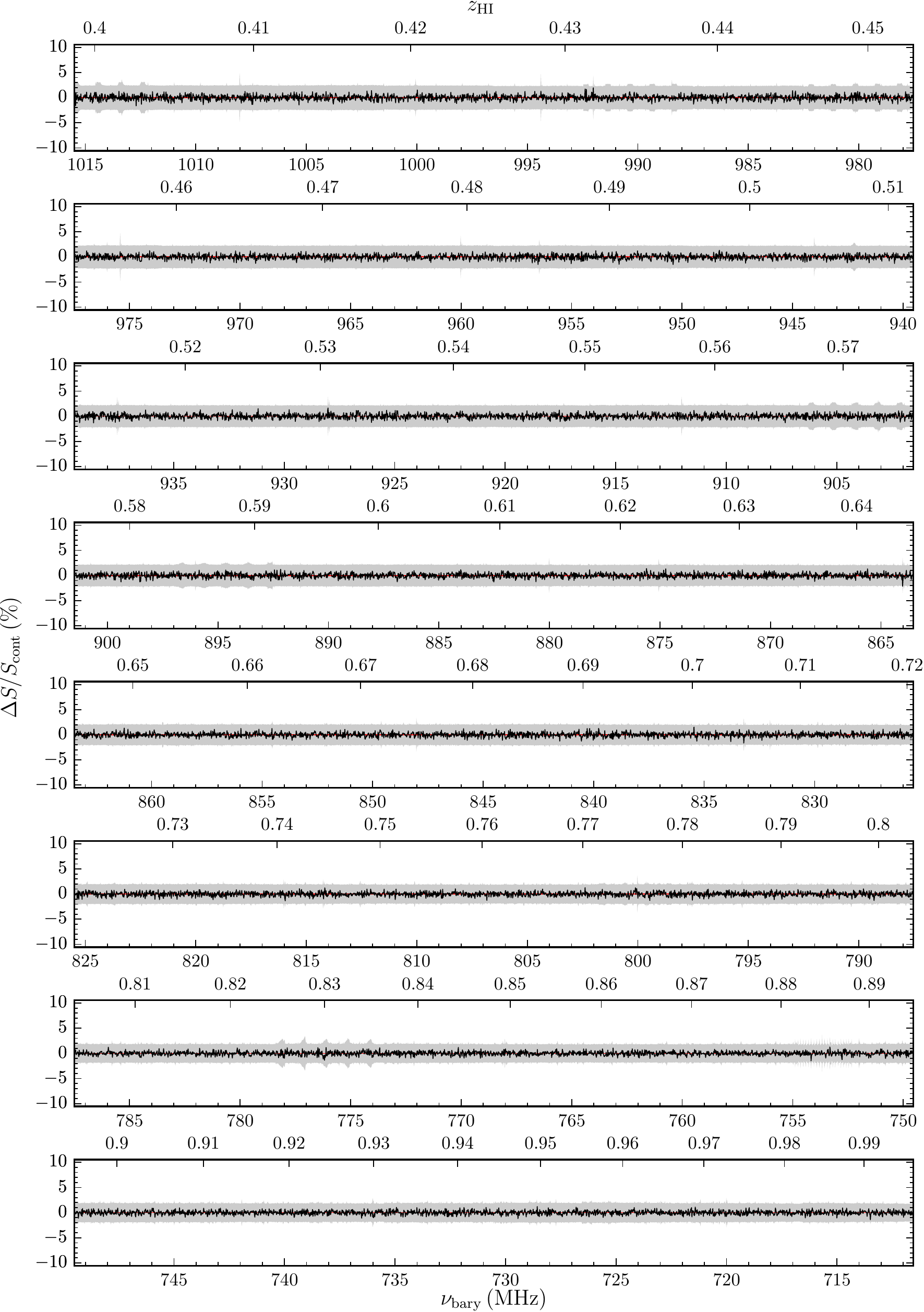}
\caption{Averaged ASKAP-BETA spectrum for PKS\,1547$-$79 based on 8\,hr of observations. The sensitivity of this averaged spectrum is 23\,mJy\,beam$^{-1}$, corresponding to a $5\sigma$ optical depth sensitivity of $\tau$ $>$ 0.019. In this figure, and each figure following, the grey region represents the 5$\sigma$ noise level in that channel.}
\label{fig:xray1547}
\end{figure*}

\begin{figure*}
\centering
\includegraphics[angle=0,width=0.8\textwidth]{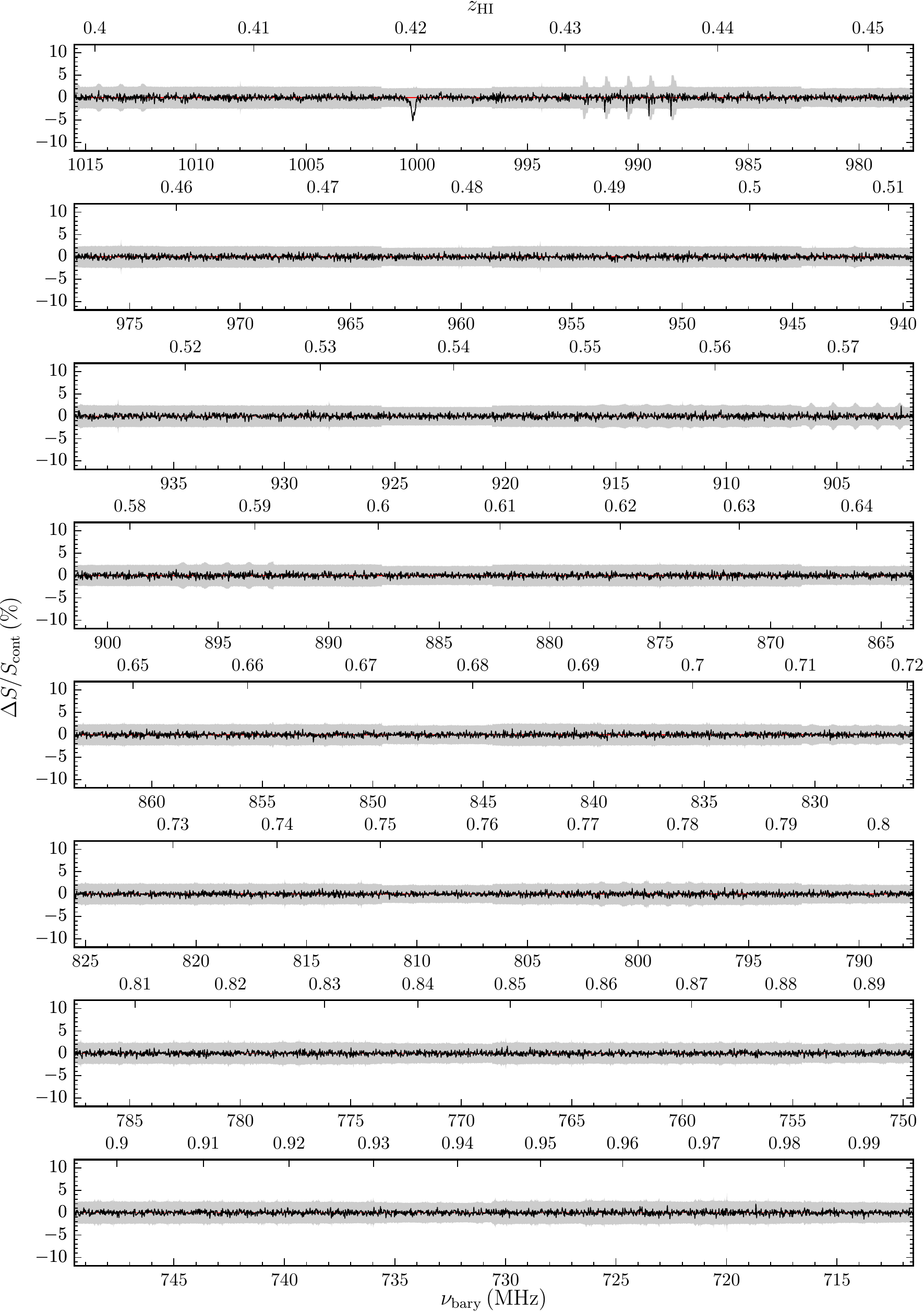}
\caption{Averaged ASKAP-BETA spectrum for PKS\,1657$-$298 based on 27.25\,hr of observations. The sensitivity of this averaged spectrum is 11\,mJy\,beam$^{-1}$, corresponding to a $5\sigma$ optical depth sensitivity of $\tau$ $>$ 0.022.}
\label{fig:xray1657}
\end{figure*}

\begin{figure*}
\centering
\includegraphics[angle=0,width=0.8\textwidth]{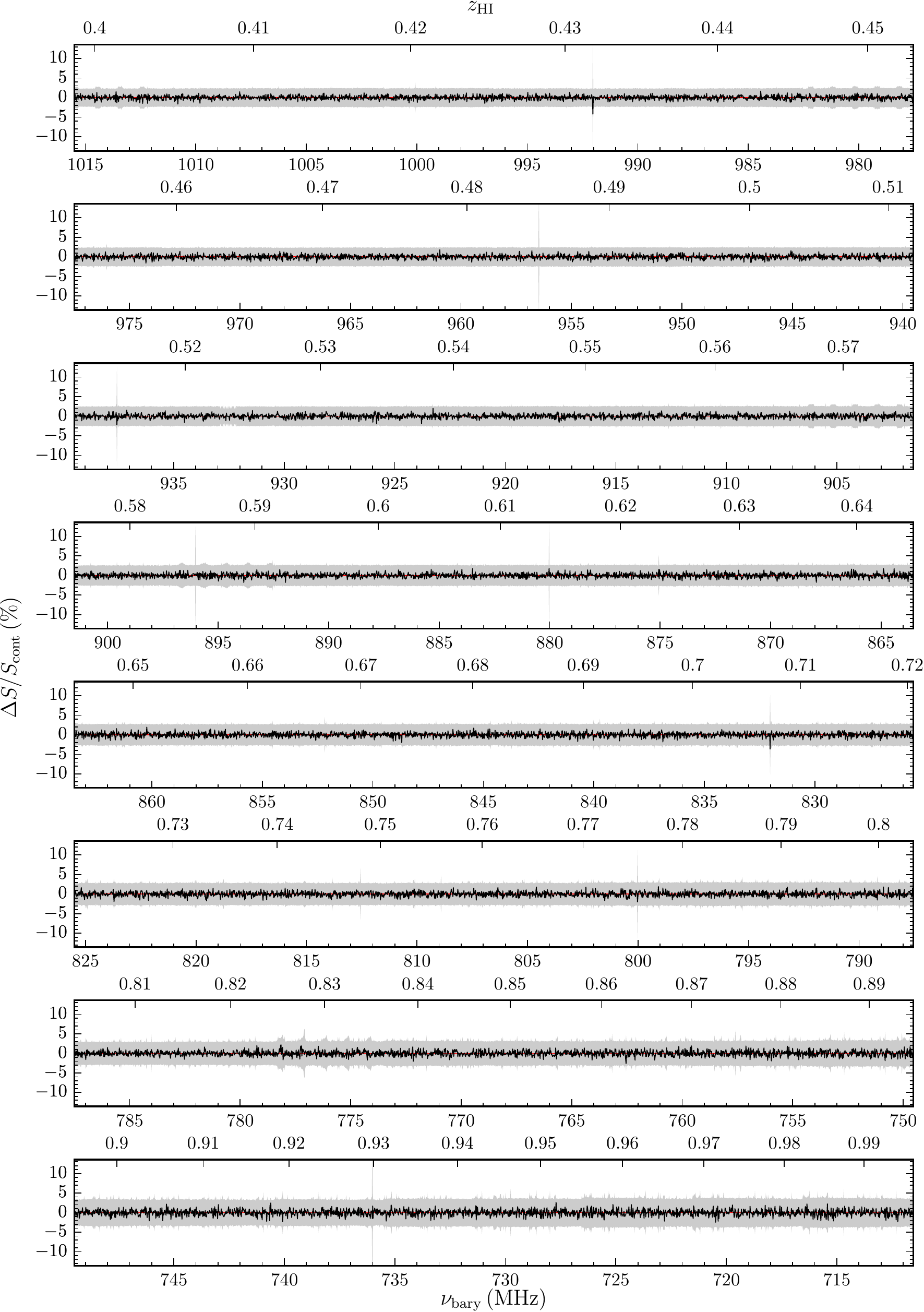}
\caption{Averaged ASKAP-BETA spectrum for MRC\,1722$-$644 based on 3\,hr of observations. The sensitivity of this averaged spectrum is 29\,mJy\,beam$^{-1}$, corresponding to a $5\sigma$ optical depth sensitivity of $\tau$ $>$ 0.032.}
\label{fig:xray1726}
\end{figure*}

\begin{figure*}
\centering
\includegraphics[angle=0,width=0.8\textwidth]{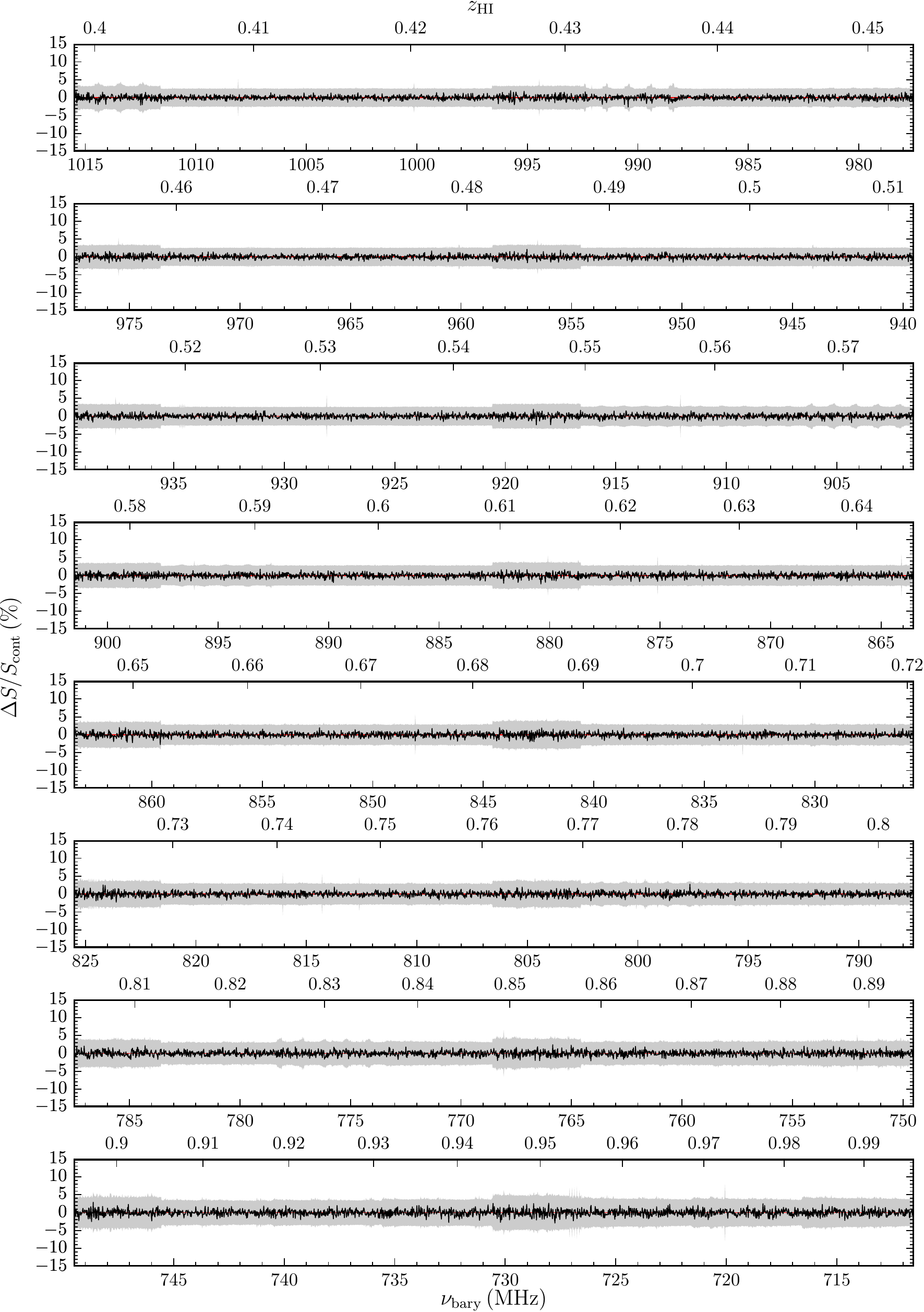}
\caption{Averaged ASKAP-BETA spectrum for PKS\,2008$-$068 based on 15.5\,hr of observations. The sensitivity of this averaged spectrum is 17\,mJy\,beam$^{-1}$, corresponding to a $5\sigma$ optical depth sensitivity of $\tau$ $>$ 0.028. An increase in noise is seen at regular intervals throughout the spectrum (every 4 chunks, e.g. at 1015\,MHz) due to correlator issues, resulting in a reduced optical depth sensitivity of $\tau$ $>$ 0.05 in this part of the spectrum. We note that the reported redshift of the host galaxy ($z$ = 0.547) does correspond to one of these affected chunks.}
\label{fig:xray2008}
\end{figure*}

\begin{figure*}
\centering
\includegraphics[angle=0,width=0.8\textwidth]{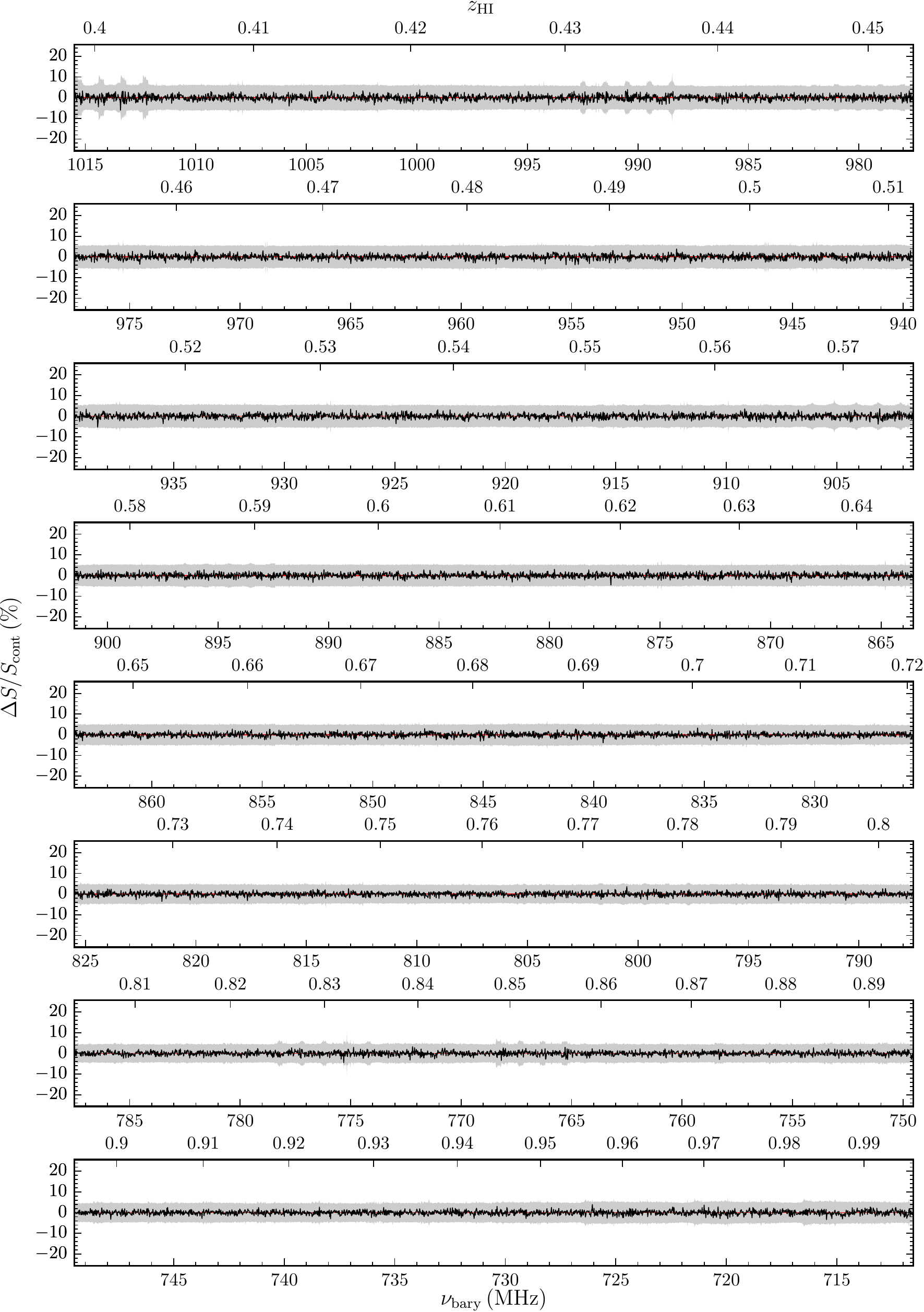}
\caption{Averaged ASKAP-BETA spectrum for PKS\,2154$-$11 based on 13.2\,hr of observations. The sensitivity of this averaged spectrum is 22\,mJy\,beam$^{-1}$, corresponding to a $5\sigma$ optical depth sensitivity of $\tau$ $>$ 0.070.}
\label{fig:xray2154}
\end{figure*}

Each source in our sample was observed with BETA for a number of hours during time available for commissioning observations. An initial 3\,hr observation was used to check for obvious absorption signatures. The minimum total integration time was chosen to ensure sensitivity to optical depths $\tau$ $\ge$ 0.05, with more time spent integrating on sources where possible. We give details of the observations in Table \ref{tb:obs}. For one source (PKS\,2154$-$11), we could only reach an optical depth sensitivity of $\tau \ge 0.07$ due to its low flux density and limits in available commissioning time. At certain times only a subset of the 6-antenna BETA was available in which case we made use of all working antennas. The data were reduced following the procedure described by \citet{2015MNRAS.453.1249A}. For an individual source, the final spectrum was produced from the combination of all good datasets available. These spectra are shown in Figures \ref{fig:xray1547} to \ref{fig:xray2154}, located at the end of this article. We identified the possible absorption signature in PKS\,1657$-$298 in the first 3\,hr observation of the source, and followed it up with several more hours of observation to obtain a higher signal to noise spectrum for analysis. We describe its properties in detail in the following section. 

\begin{figure}
\centering
\includegraphics[angle=0,width=0.45\textwidth]{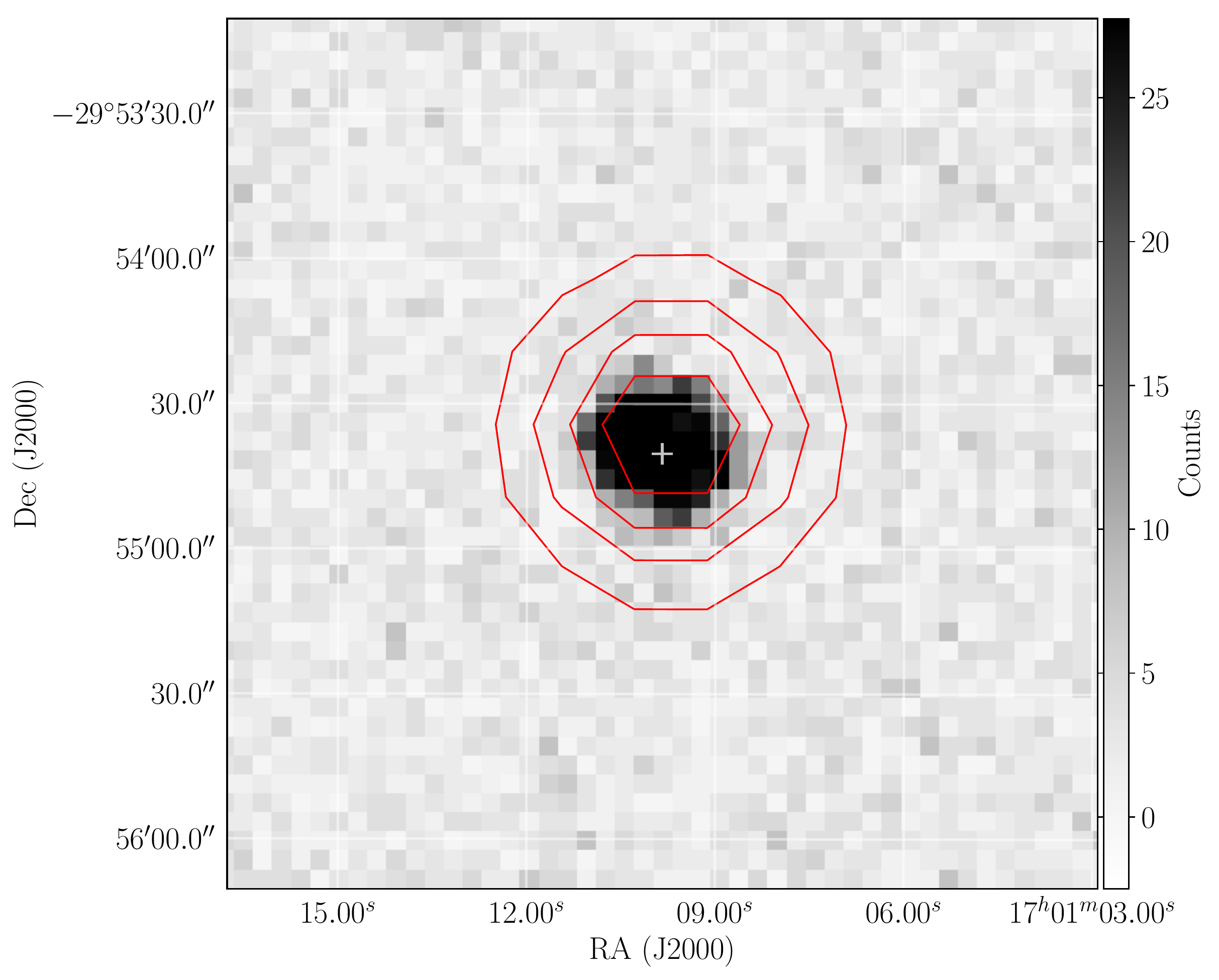}
\caption{X-ray image of PKS\,1657$-$298 from {\it Chandra} data (obtained via the {\it Chandra} Search and Retrieval service for observation ID 13711), overlaid with radio contours from NVSS at 1.4\,GHz (shown in red). The contours range from 20\% to 80\% of the peak in 4 equally-spaced levels.}
\label{fig:radioxray}
\end{figure}

\begin{table*}
\centering
\caption{Table containing details of the observations of each galaxy in our BETA sample. We record here the date of the observation, the total hours spent integrating on source, the antennas included in the array, the resulting RMS noise per channel and corresponding 5$\sigma$ optical depth sensitivity.}\label{tb:obs}
\begin{tabular}{ccccccccc}
\hline
Source & Date & $t_{int}$ & Antenna numbers & $\sigma_{chan}$ & 5$\sigma_\tau$\\ 
&&hr&&mJy beam$^{-1}$&&\\
\hline
PKS\,1547$-$79	
& 14 Nov 2014 & 3 & 1,\,3,\,6,\,8,\,9,\,15 & 29 & 0.024\\
& 29 Jan 2016 & 5 & 1,\,3,\,8,\,15 & 35 & 0.029 \\
&Total & 8 & & 23 & 0.019 \\
\hline
PKS\,1657$-$298	
& 21 Nov 2014 & 3 & 1,\,3,\,6,\,8,\,9,\,15 & 29 & 0.058\\
& 5 May 2015 & 3.25 & 1,\,3,\,6,\,8,\,9,\,15 & 28 & 0.056 \\
& 9 Jun 2015 & 10 & 1,\,3,\,6,\,8,\,15 & 19 & 0.038\\
& 24 Aug 2015 & 6 & 1,\,3,\,6,\,8,\,15 & 25 & 0.050\\
& 02 Oct 2015 & 5 & 1,\,3,\,6,\,8,\,15 & 27 & 0.054\\
&Total & 27.25 & & 11 & 0.022\\
\hline
MRC~1722$-$644	
& 15 Nov 2014 & 3 & 1,\,3,\,6,\,8,\,9,\,15 & 29 & 0.032\\
&Total & 3 & & 29 & 0.032\\
\hline
PKS\,2008$-$068
& 22 Nov 2014 & 2 & 1,\,3,\,6,\,8,\,9,\,15 & 35 & 0.058\\
& 23 Nov 2014 & 2 & 1,\,3,\,6,\,8,\,9,\,15 & 35 & 0.058\\
& 9 Jun 2015 & 3 & 1,\,3,\,6,\,8,\,15 & 35 & 0.058\\
& 5 Oct 2015 & 5 & 1,\,3,\,6,\,8,\,15 & 27 & 0.045\\
& 30 Jan 2016 & 1.5 & 1,\,3,\,8,\,15 & 64 & 0.107\\
& 1 Feb 2016 & 2 & 1,\,3,\,8,\,15 & 55 & 0.092\\
& Total & 15.5 & & 17 & 0.028 \\
\hline
PKS\,2154$-$11
& 7 May 2015 & 3 & 1,\,3,\,6,\,8,\,9,\,15 & 29 & 0.097\\
& 22 Oct 2015 & 3.1 & 1,\,3,\,6,\,8,\,15 & 35 & 0.117 \\
& 27 Jan 2016 & 1.2 & 1,\,3,\,8,\,15 & 72 & 0.240\\
& 27 Jan 2016 & 0.9 & 1,\,3,\,8,\,15 & 83 & 0.277\\
& 27 Jan 2016 & 2 & 1,\,3,\,8,\,15 & 55 & 0.183\\
& 31 Jan 2016 & 3 & 1,\,3,\,8,\,15 & 45 & 0.150 \\
& Total & 13.2 & & 22 & 0.070\\
\hline
\end{tabular}
\end{table*}

\section{PKS\,1657$-$298: a new H{\sc i} absorber}
From our BETA pilot study, we discovered a new H{\sc i} absorption line towards PKS\,1657$-$298. Of the five sources observed, this radio source featured the largest X-ray hardness ratio of 424 (whereas the others were closer to $\sim$100), thus making this detection of H{\sc i} absorption consistent with our findings in Section \ref{propcompare}. As described in Section \ref{betaprop}, there is very little known about this source in general, most probably due to its inconvenient location at a Galactic position of $l$ = 353.7$^\circ$, $b$ = 7.4$^\circ$, making it an unlikely target for optical and infrared studies. Figure \ref{fig:radioxray} shows an X-ray image from {\it Chandra} (obtained from the online archive) of PKS\,1657$-$298 overlaid with radio contours from NVSS at 1.4\,GHz. There are several serendipitous observations of this source in both the {\it Chandra} and {\it XMM-Newton} archives due to its spatial location near to the low mass X-ray binary MXB\,1659$-$298. 

As stated in Section \ref{betaprop}, there is no known redshift for PKS\,1657$-$298. Aside from the H{\sc i} detection at a redshift of $z$ = 0.42, we do not currently have other multi-wavelength evidence confirming the same redshift for our assumed host radio galaxy. There is also the possibility that the coincident radio and X-ray sources are not physically associated. To address this, if we consider the number density of X-ray sources in the 3XMM-DR4 release equal to or brighter than the integrated flux of PKS\,1657$-$298 (14\,deg$^{-2}$ for fluxes $\ge$ 7.5$\times$10$^{-13}$\,erg\,cm$^{-2}$\,s$^{-1}$) and the number density of radio sources in NVSS similarly equivalent in flux density (0.02\,deg$^{-2}$ for flux densities $\ge$ 1777\,mJy), then the Poisson probability of finding a source of each kind within a 30$''$ radius is 0.000001\%, driven mostly by the radio brightness of PKS\,1657$-$298. If we instead consider that we specifically selected bright radio sources and ask what the chance of any X-ray source nearby is (based on the density of 3XMM-DR4 sources of 469\,deg$^{-2}$), then we find a similarly-low probability of 0.004\%. As such, based on this alone we conclude that the radio and X-ray sources are almost certainly physically associated. In order to detect the H{\sc i} absorption, we know that the redshift of the radio source is definitely $z$ $\ge$ 0.42 as it must be acting as a background source. We can thus consider the implications of the well-studied $K$-$z$ relation for radio/infrared galaxies \citep{Lilly:1984ke}, which describes the correlation between $K$-band magnitude (2.17\,$\mu$m) and redshift $z$. In particular, we compare the infrared properties of PKS\,1657$-$298 with the results of \citet{Willott:2003ep}. Although the galaxy is visible in 2MASS-K, it is not catalogued, possibly due to its proximity to a nearby source of similar brightness (17011004$-$2954423). We applied basic aperture photometry to estimate the individual brightnesses of each source, and we find that they are consistent within errors to each other. Thus, we can adopt the same magnitude for PKS\,1657$-$298, indicating a $K$-band magnitude of 14.9$\pm$0.1 as catalogued in 2MASS. There is a chance that the recorded magnitude is due to the blended combination of each source, in which case the true $K$-band magnitude would be $\sim$15.6. However, when applying the same basic aperture photometry to nearby objects of similar counts (namely, 17011007$-$2955021 and 17011032$-$2954228), we find similar $K$-band magnitudes of $\sim$14.9 which suggests that this is an accurate estimate for PKS\,1657$-$298 also. Nonetheless, we also consider the `deblended' scenario below. 

\begin{figure}
\centering
\includegraphics[angle=0,width=0.45\textwidth]{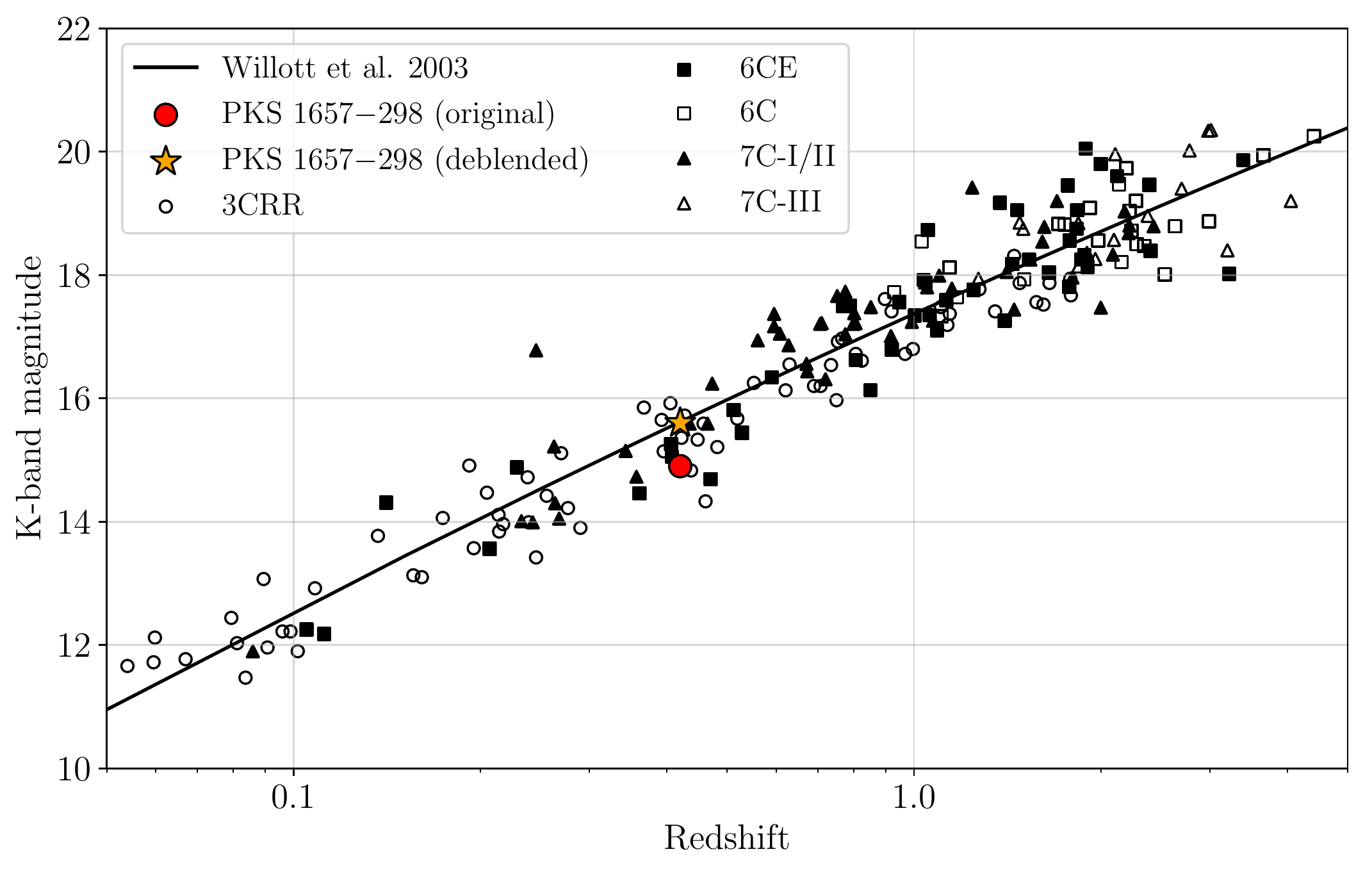}
\caption{Scatter plot of $K$-$z$ relation data from \citet{Willott:2003ep}, including their best fit line to the data. We have plotted the position of PKS\,1657$-$298 as a red circle, based on its redshift assumed from the H{\sc i} absorption data and $K$-band magnitude from 2MASS. The deblended magnitude assuming blending with 17011004$-$2954423 is shown as an orange star. We see that this is in good agreement with the $K$-$z$ relation, and in fact the assumption of a higher redshift for the galaxy is less consistent with existing data.}
\label{fig:kzrel}
\end{figure}

Assuming the same redshift as derived from the H{\sc i} absorption, we plot the $K$-$z$ position of PKS\,1657$-$298 (red circle) in Figure \ref{fig:kzrel} alongside the other datasets as studied in \citet{Willott:2003ep}. Assuming the 2MASS-K magnitude is blended with the nearby source, we additionally plot the deblended magnitude for PKS\,1657$-$298 (orange star) for comparison. We also plot the best fit reported in \citet{Willott:2003ep} as a solid line. Based on the $K$-$z$ relation data, we see that a redshift of $z$ = 0.42 is very much consistent with the 2\,$\mu$m properties of PKS\,1657$-$298 (either catalogued or deblended) and that a higher redshift (that is, a more distant radio galaxy) is actually less consistent with the relation. We note that reddening due to Galactic extinction ($\sim$0.1\,mag at 2\,$\mu$m) has very little impact on this result and is well within the scatter, and further that, if not blended with 17011004$-$2954423, PKS\,1657$-$298 is slightly brighter than we might expect based on the $K$-$z$ relation which may indicate an unusually luminous host. Based on this result, we conclude that the H{\sc i} absorption detected towards PKS\,1657$-$298 is likely to be located at the same redshift as the coincident radio/X-ray sources. 

\begin{figure*}
\centering
\includegraphics[angle=0,width=0.45\textwidth]{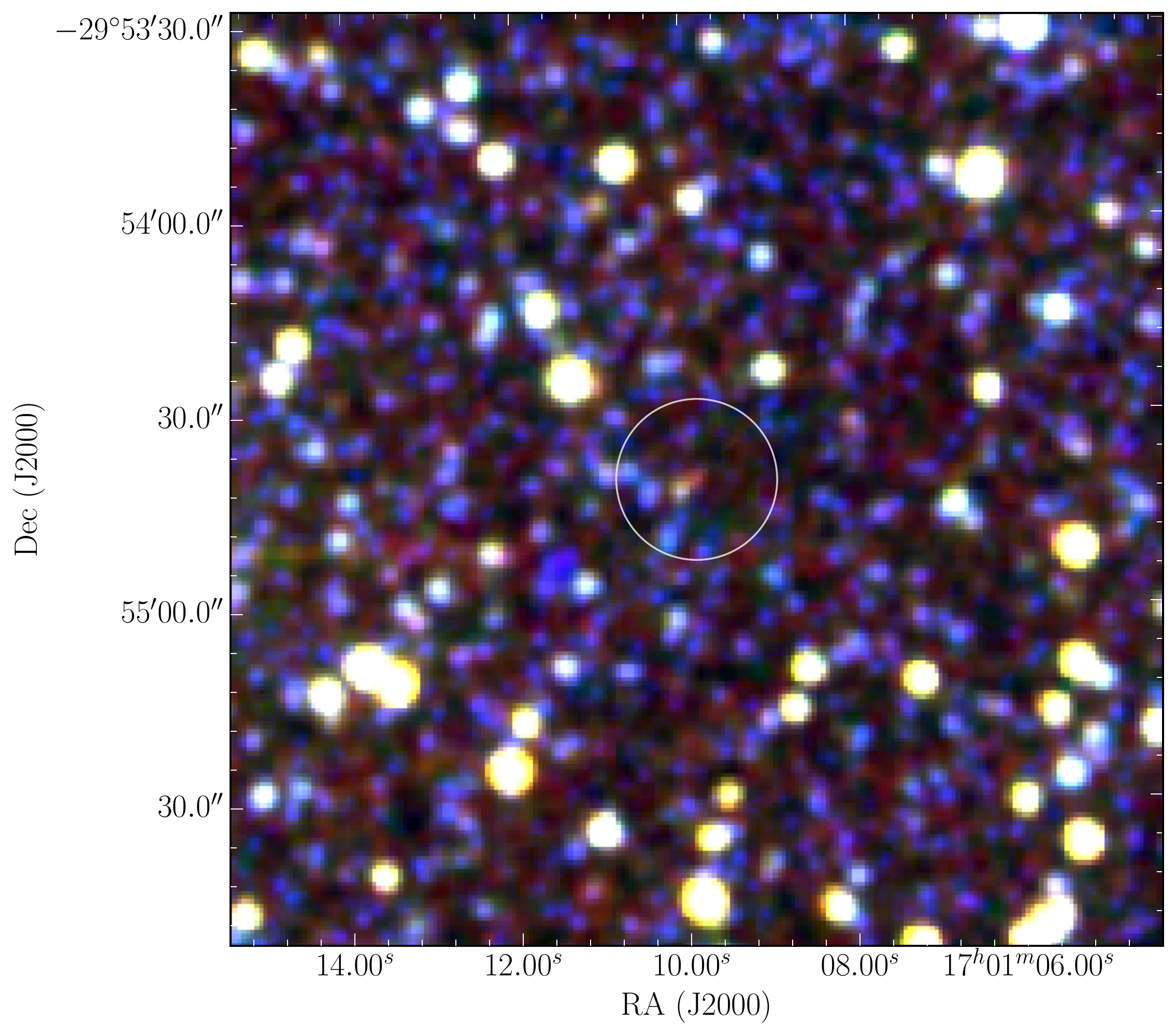}
\includegraphics[angle=0,width=0.45\textwidth]{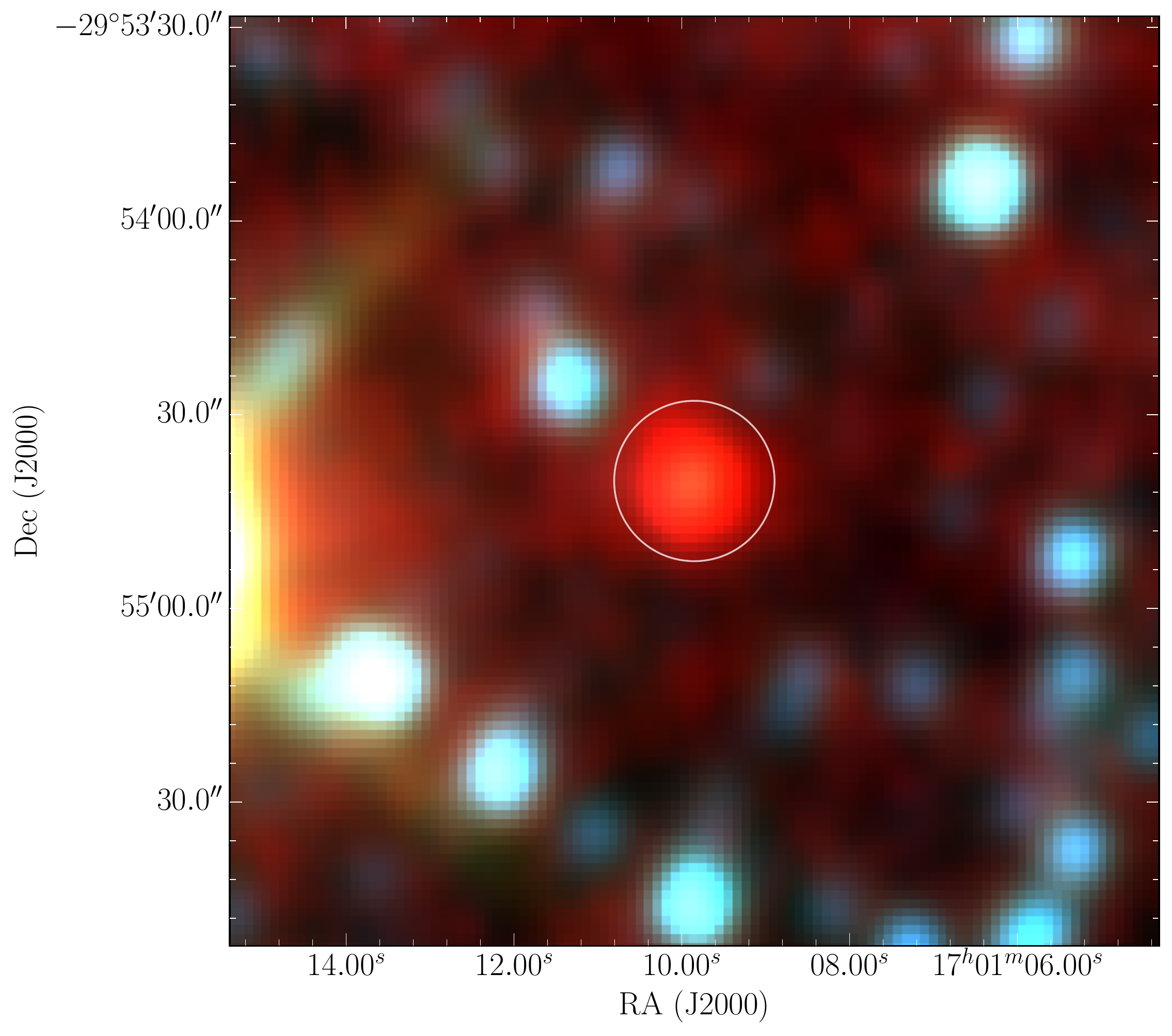}
\caption{Three-colour optical/infrared images of PKS\,1657$-$298. Left panel: image formed from 2MASS-K (red), 2MASS-J (green) and UKST-B (blue) with each image scaled to 95\% percentile. A small red blob is seen at 2\,$\mu$m, with non-detections in the other bands. Right panel: image formed from WISE 12\,$\mu$m (red), 4\,$\mu$m (green) and 3\,$\mu$m (blue) also scaled to their 95th percentiles. The galaxy is detected clearly in each band but is brightest at 12\,$\mu$m. In both cases, the white circles are centred on the VLBI position measured by \citet{Petrov:2012if}.}
\label{fig:1657rgb}
\end{figure*}

We conducted a multi-wavelength search in order to characterise the properties of this source as best as possible based on existing data. Figure \ref{fig:1657rgb} shows three-colour images of PKS\,1657$-$298 using available optical/infrared data from DSS/DSS2 \citep{Lasker:1990gh,McLean:2000vf}, SuperCOSMOS \citep{Hambly:2001dq}, 2MASS \citep{Skrutskie:2006hl} and WISE \citep{Wright:2010in}. The left panel is formed from 2MASS-K (red), 2MASS-J (green) and UKST-B (blue) with each image scaled to 95\% percentile. There is a small source evident in the 2MASS-K image at 2\,$\mu$m, although it is absent from the other two bands and also not detected in 2MASS-J or 2MASS-H. This absence of detection cannot be solely attributed to Galactic extinction, given that 2MASS-J and 2MASS-H are more sensitive than 2MASS-K and that it is very tentatively visible in ESO-R (implying an $R$-band magnitude $\gtrsim$ 21 mag). Instead, the lack of clear detection at optical/infrared wavelengths (with the exception of $K$-band) implies a high level of intrinsic reddening due to dust and gas. This is consistent with the recent findings of \citet{Curran:2017bk}, specifically that the detection rate of H{\sc i} absorption increases towards 100\% as the $B$-$K$ colour increases. The right panel of Figure \ref{fig:1657rgb} shows a three-colour image formed from WISE 12\,$\mu$m (red), 4\,$\mu$m (green) and 3\,$\mu$m (blue) also scaled to their 95th percentiles in intensity. PKS\,1657$-$298 is detected significantly in all WISE bands, however it has a considerable excess in emission at 12\,$\mu$m causing the evident reddening in the image. This is consistent with the conclusion based on DSS/SuperCOSMOS/2MASS that it is intrinsically (strongly) reddened. The overlaid white circles in each panel are centred on the VLBI position for PKS\,1657$-$298 from \citet{Petrov:2012if}. As seen in Figure \ref{fig:wiseplot}, PKS\,1657$-$298 occupies a region of the diagram shared by Seyferts and LIRGs which gives us the current best insight into the properties of the host without better optical data. 

The VLBI data currently available for PKS\,1657$-$298 are limited, however do provide further insight into the properties of the host. From previous observations of the source at 2.3\,GHz and 8.4\,GHz (accessed via http://astrogeo.org), we can estimate the diameter of the radio source (apparently split into two compact components) to be between $\sim$10--15\,mas. At a redshift of $z$ = 0.42, this corresponds to a physical extent of $\sim$50--80\,pc which is consistent with a typical compact symmetric object (CSO). Based on the limited uv-coverage and sensitivity of these current data, we do not have information about whether we are seeing two jet components or a core/jet system, nor can we determine the relative flux density of each component. 

\subsection{H{\sc i} absorption properties}\label{betadata}
\begin{figure}
\centering
\includegraphics[angle=0,width=0.45\textwidth]{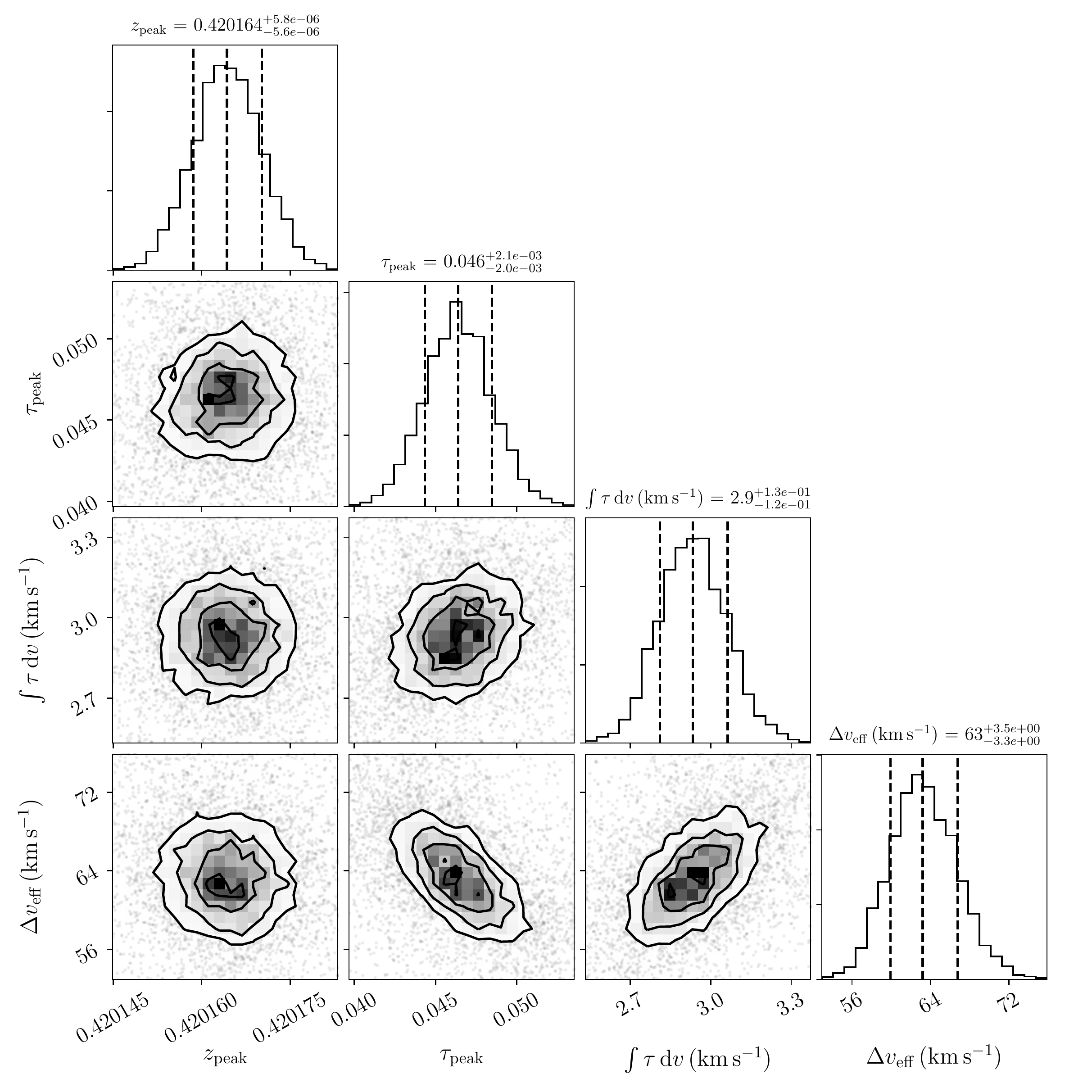}
\caption{Marginal posteriors for the properties of H{\sc i} absorption towards PKS\,1657$-$298. From these, we measure the redshift $z_{\rm peak}$, peak optical depth $\tau_{\rm peak}$, integrated optical depth $\int \tau {\rm d}v$ $({\rm km\,s}^{-1})$, and effective line-width $\Delta v_{\rm eff}$ $({\rm km\,s}^{-1})$.}
\label{fig:1657dists}
\end{figure}

We describe here our analysis of the H{\sc i} absorption towards PKS\,1657$-$298 as detected with BETA. We model the absorption line as a superposition of Gaussian profiles and then determine the best fitting number of components and marginal distributions for the line parameters using a Bayesian approach as described by \citet{2015MNRAS.453.1249A,Allison:2016gd} and references therein. We identify a single component in the absorption-line profile of PKS\,1657$-$298, at a redshift of $z = 0.420164 \pm 0.000006$. We show the marginal posteriors for key parameters in Figure \ref{fig:1657dists}. The peak and integrated optical depths of the line are $0.046 \pm 0.002$ and $\int \tau {\rm d}v$\,km~s$^{-1}$ = $2.9 \pm 0.1$, respectively. Dividing these two values we obtain an effective line-width of 63$^{+4}_{-3}$\,km~s$^{-1}$. Assuming a spin temperature of 100\,K and a covering factor of 1 (to maintain consistency with previous work), we obtain a column density of $N_{\rm HI}$ = 5.3 $\pm$ 0.3 $\times$ 10$^{20}$\,cm$^{-2}$. 

\begin{figure}
\centering
\includegraphics[angle=0,width=0.45\textwidth]{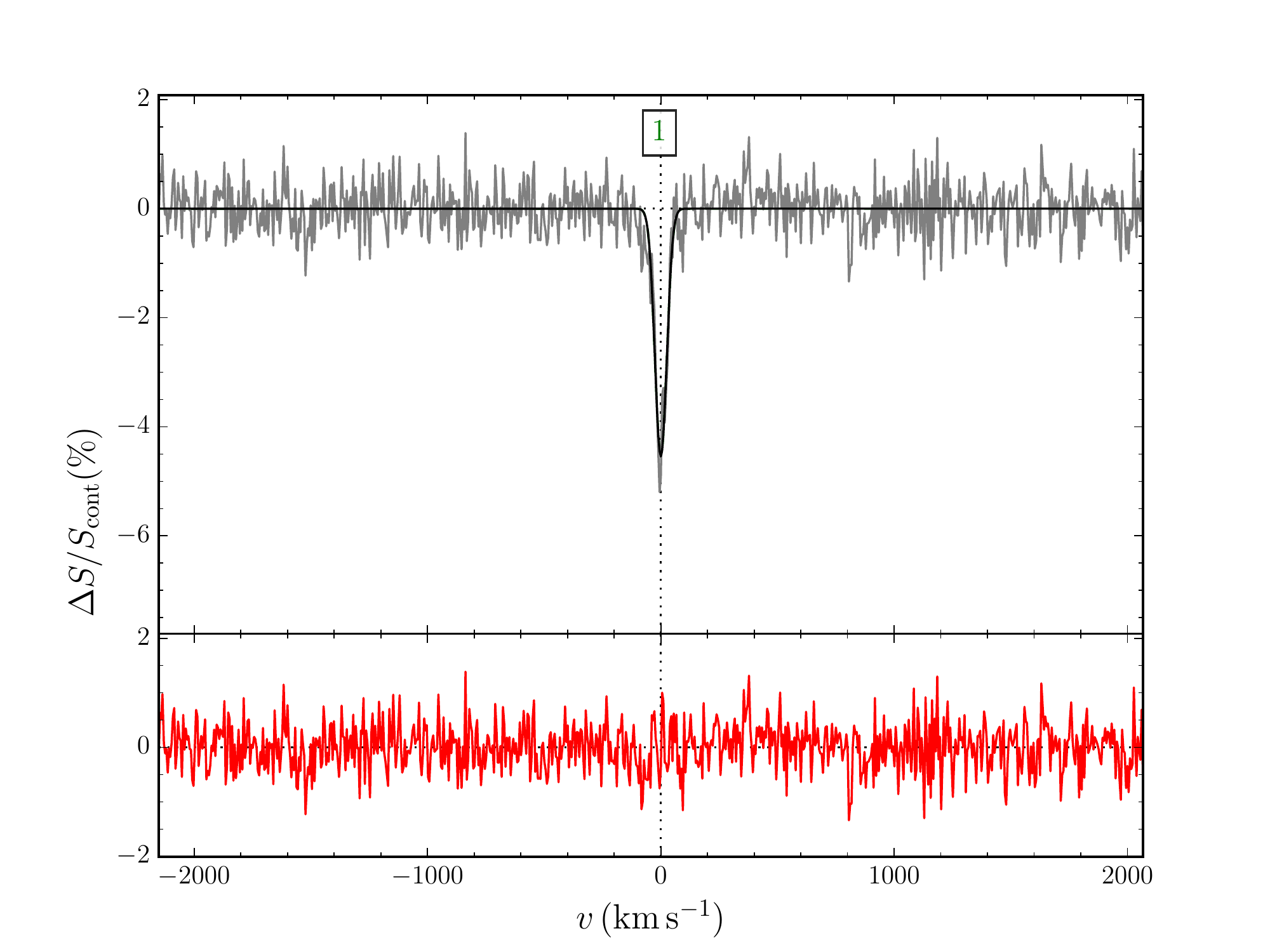}
\caption{The best fitting model (black) for our average PKS\,1657$-$298 spectrum (grey). The residual spectrum is shown in red. Details of our method are given in Section \ref{betadata}.}
\label{fig:1657line}
\end{figure}

These properties are consistent with those typically measured for H{\sc i} absorption associated with radio AGN \citep[e.g.][]{Gereb:2015cx}. The symmetry and width of the line are consistent with the compact radio source illuminating a regular gaseous structure (such as a disk) near to the systemic velocity of the host galaxy. There is some hint at low significance of structure in the main line (potentially two components instead of one) as well as a possible extension bluewards of the main line as seen in Figure \ref{fig:1657line}. This may be evidence of individual dense clouds or broad outflows. However, as these are below statistical significance in our current data, we defer further speculation to future radio observations of PKS\,1657$-$298. Of key interest from this point is comparing our properties as measured from the H{\sc i} absorption with those indicated by the X-ray data available to us for this galaxy.

\subsection{X-ray properties: {\it Chandra}}
\begin{figure}
    \centering
    \includegraphics[angle=270,width=0.46\textwidth]{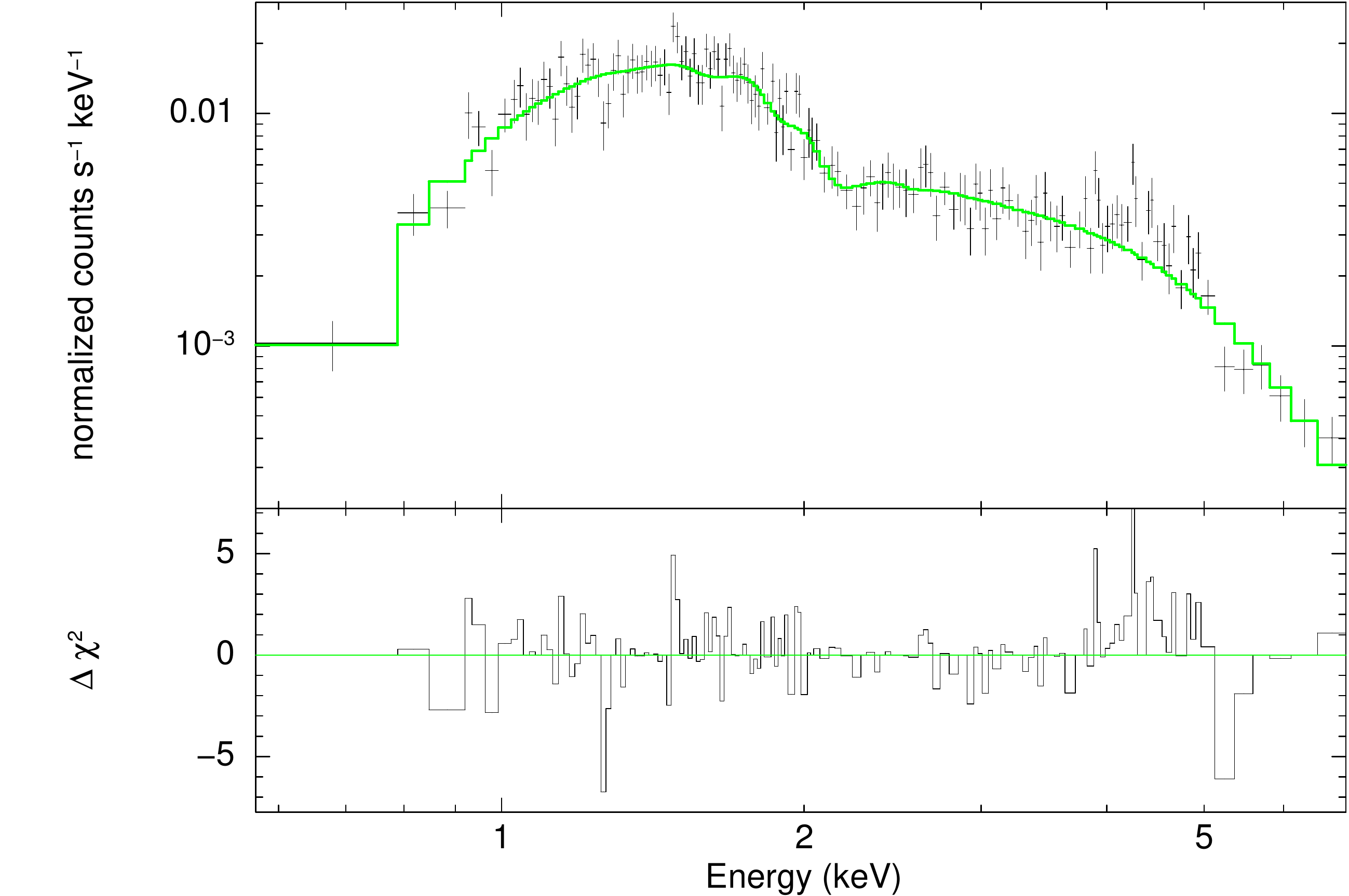}
    \includegraphics[angle=270,width=0.46\textwidth]{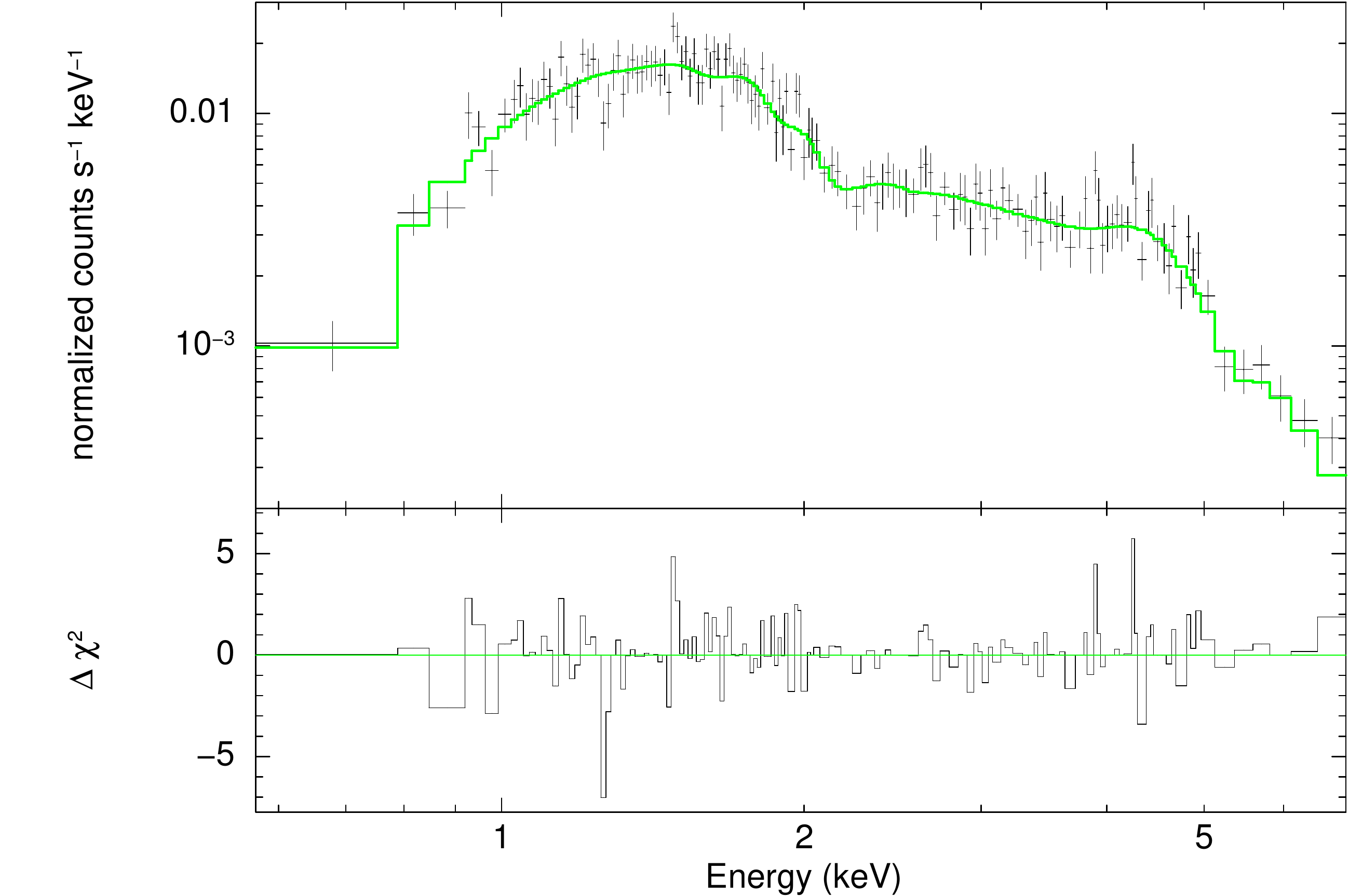}
    \caption{Top: Stacked {\it Chandra} spectrum for PKS\,1657$-$298 with model fits and residuals. A simple absorbed power-law is used ({\it tbgrain $\times$ ztbabs $\times$ powerlaw}). This results in significant high-energy residuals. Bottom: As in above but we introduce two Gaussian components (one with a negative normalisation) to account for the residuals ({\it tbgrain $\times$ ztbabs $\times$ (gauss + gauss+ powerlaw)}). Both spectra have been rebinned to a signal-to-noise of at least 4 for clarity.}
    \label{chan_spec_fig}
    \vspace{0.3cm}
\end{figure}

{\it Chandra} observations including PKS\,1657$-$298 in the field of view (observation IDs 5469, 6337, 13711, and 14453) were combined into a single spectrum. The stacked spectrum was then fitted with a simple absorbed power-law model ({\it tbgrain$\times$ztbabs$\times$powerlaw}). We adopted these models instead of the standard {\it phabs} and {\it zphabs} as they feature more precise elemental cross-sections and allow us to incorporate the presence of line-of-sight molecular hydrogen, which is important at such low Galactic latitudes. In this model, {\it tbgrain} accounts for the line-of-sight absorption and {\it ztbabs} for the intrinsic absorption at the redshift of the source (see \citealt{Wilms:2000en} for details of the photo-ionisation cross-sections in the Tuebingen-Boulder ISM absorption models). We adopted a value of $N_{\rm H,Gal}$ = 3.1 $\times$ 10$^{21}$\,cm$^{-2}$ for the line-of-sight component, which is the sum of $N_{\rm HI,Gal}$ = 1.8 $\times$ 10$^{21}$\,cm$^{-2}$ and $N_{\rm H2,Gal}$ = 1.3 $\times$ 10$^{21}$\,cm$^{-2}$ using the method of \citet{Willingale:2013kz}. This was used as an input parameter to {\it tbgrain} (with default values for the grain phase), where we find that the total column is not very sensitive to changes to the molecular hydrogen fraction. The power-law index returned from this model was $\Gamma = 1.84$~$\pm$~$0.10$. This corresponded to a decent ($\chi_{\nu}^2 \approx 1.12$) fit, however with high-energy residuals present (Figure \ref{chan_spec_fig}, top).

We added two Gaussian components to our spectral model: one to simulate an emission line and the other (with a negative normalisation) an absorption edge (Figure \ref{chan_spec_fig}, bottom). Here, the power-law index of the continuum is found to be $\Gamma=1.93\pm0.15$. We varied the emission line energy between 4 and 6\,keV, finding a best fit of $E_{\rm em} = 4.37^{+0.18}_{-0.17}\,$keV and equivalent width of $\sigma_{\rm em} = 0.33^{+0.23}_{-0.16}\,$keV. At those energies, the intrinsic energy resolution (observed FWHM of a narrow line) of the ACIS-S3 chip\footnote{http://cxc.cfa.harvard.edu/proposer/POG/html/} is $\sim 0.16$\,keV. The absorption edge energy was tested from 4.5 to 6.5\,keV, with a best fit of $E_{\rm abs} = 5.4\pm 0.4$\,keV. This improved the fit to $\chi_{\nu}^2 \sim 1.02$, with an F-test significance of $99.9\%$. The emission line can best be interpreted as a narrow Fe line at $\sim$6.4\,keV in the rest frame of PKS\,1657$-$298; however, we cannot rule that we are instead looking at a broader Fe line at $\sim$6.7\,keV. The absorption line is consistent with an Fe K-shell absorption edge at a rest-frame energy of 7.1\,keV, which is expected from Fe in the same low-ionisation states that produce the 6.4 keV emission line. Therefore, the observed energies of these features indicate a redshift of $\sim$0.42, consistent with the redshift estimated from the H{\sc i} data. Alternatively, we cannot rule out the possibility that both the emission line and absorption edge could be replaced by a P-Cygni profile. This scenario corresponds to a strong outflow as seen in high Eddington-ratio quasars such as PDS\,456 \citep{2015Sci...347..860N} and PG1211+143 \citep{2009MNRAS.397..249P}. Based on these {\it Chandra} data, we find an observed 0.3--10.0\,keV flux $f_{\rm x} = 4.9^{+0.3}_{-0.3}\times10^{-13}$ erg\,cm$^{-2}$\,s$^{-1}$ in the observer's frame, and we infer an unabsorbed 0.3--10.0\,keV rest frame luminosity (at $z=0.42$) of $L_{\rm x} = 5.2^{+0.4}_{-0.3}\times10^{44}$\,erg\,s$^{-1}$. We derive a total hydrogen column towards PKS\,1657$-$298 of $N_{\mathrm{H,tot}}$ = $0.9^{+0.2}_{-0.2} \times 10^{22}$\,cm$^{-2}$, of which we estimate that $N_{\mathrm{H,X}}$ = $0.6^{+0.2}_{-0.2} \times 10^{22}$\,cm$^{-2}$ is intrinsic to the emitting source.

\subsection{X-ray properties: {\it XMM-Newton}}
\begin{figure}
    \centering
    \includegraphics[angle=270,width=0.46\textwidth]{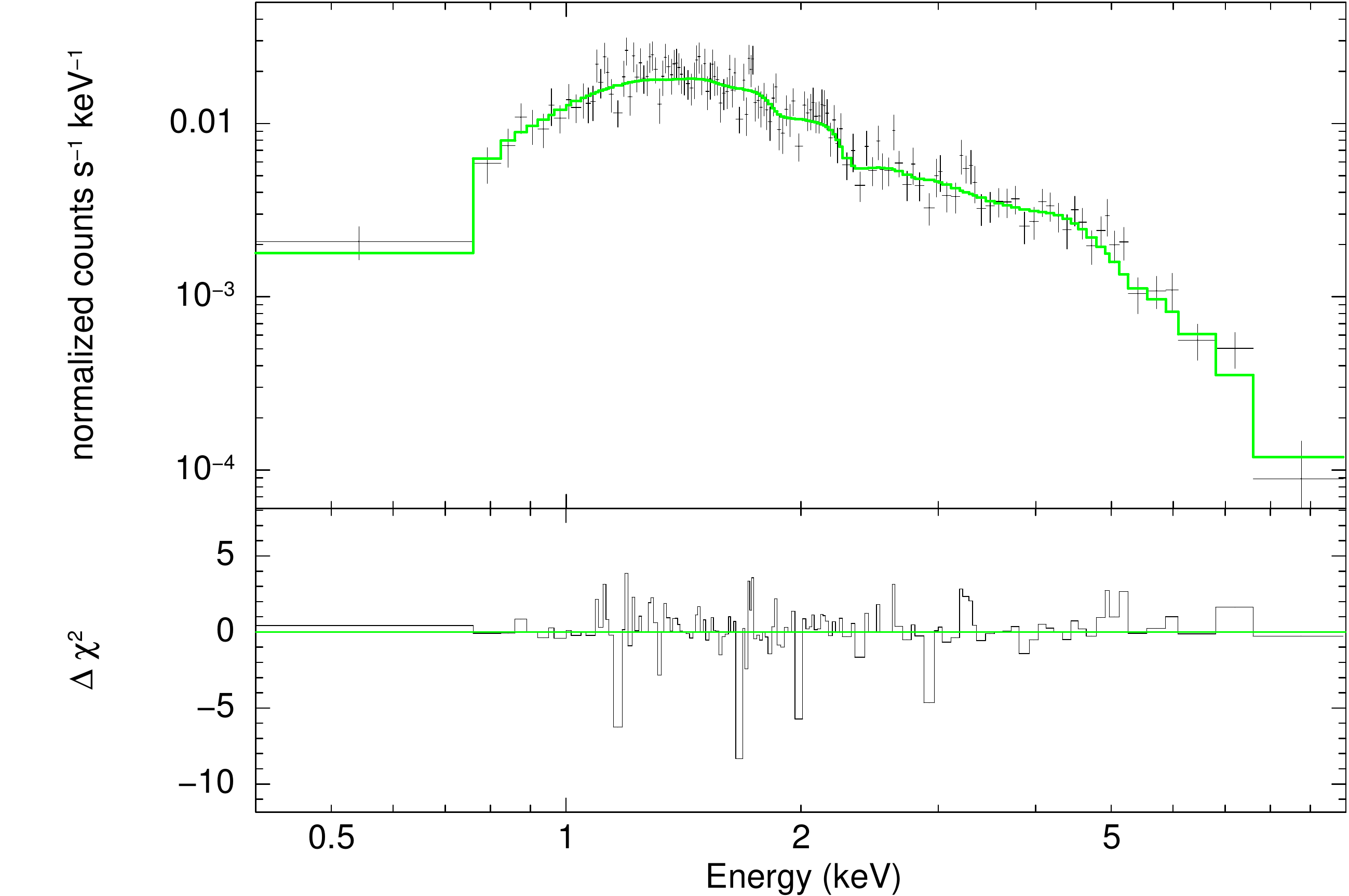}
    \caption{As in Figure \ref{chan_spec_fig} but for the stacked {\it XMM-Newton} spectrum. Once again, the spectrum has been rebinned to a minimum signal-to-noise of 4.}
    \label{xmm_spec_fig}
    \vspace{0.3cm}
\end{figure}

 We also used archival {\it XMM-Newton} data of PKS\,1657$-$298. We combined the EPIC pn and MOS spectra (and associated response files) from observations 0153190101 and 0205580201 and the EPIC MOS spectra from observations 0008620701 and 0748391601 (the EPIC pn spectra were not available for these observations) into a single stacked spectrum in order to increase the signal-to-noise. As we did for the combined {\it Chandra} spectrum, we initially fitted the source with a simple absorbed power-law. This resulted in an acceptable fit ($\chi_{\nu}^2 \approx 1.09$). Unlike with {\it Chandra}, the signal-to-noise in the stacked spectrum is not high enough to significantly identify residual components, and the addition of a narrow emission line and an absorption edge, or of an emission plus absorption line (to match the {\it Chandra} spectrum) does not improve the fit statistics ($\chi_{\nu}^2 \approx 1.09$). The 90\%-confidence-level upper limit of the normalisation of a possible narrow emission line is $K \approx 6.7\times10^{-6}$. This is consistent with the equivalent width of the line detected in the {\it Chandra} spectrum. The continuum plus emission and absorption line model can be seen in Figure \ref{xmm_spec_fig}. The power-law index derived is $\Gamma = 1.89\pm0.14$, consistent with the {\it Chandra} spectrum, as is the observed 0.3--10.0\,keV flux and unabsorbed rest frame luminosity; $f_{\rm x} = 4.8^{+0.3}_{-0.3} \times 10^{-13}$\,erg\,cm$^{-2}$\,s$^{-1}$, $L_{\rm x} = 4.9^{+0.3}_{-0.3} \times10^{44}$\,erg\,s$^{-1}$. Also in agreement with our above analysis of available {\it Chandra} data, we derive a total hydrogen column of $N_{\mathrm{H,tot}}$ = $0.8^{+0.3}_{-0.3} \times 10^{22}$\,cm$^{-2}$ based on the {\it XMM-Newton} spectrum, of which $N_{\mathrm{H,X}}$ = $0.5^{+0.2}_{-0.1} \times 10^{22}$\,cm$^{-2}$ is estimated to be intrinsic to PKS\,1657$-$298.
 
\subsection{Constraints from radio and X-rays}
We consider here what constraint we can place on the physical properties of the galaxy based on the combination of our radio and X-ray data. To investigate the spin temperature, we take the same approach here as described in \citet{2015MNRAS.453.1249A}, such that we can find an upper limit on spin temperature, 

\begin{equation}
T_s \lesssim 5485~\Big[\frac{N_{\mathrm{H,X}}}{10^{22}\,\mathrm{cm^{-2}}}\Big]~\Big[\frac{\int \tau_{21} v\,\mathrm{d}v}{1\,\mathrm{km\,s^{-1}}}\Big]^{-1}\,\mathrm{K},
\end{equation}
~\\
where $T_s$ is the spin temperature in K, $N_{\mathrm{H,X}}$ is the total hydrogen column density as measured from the X-rays in cm$^{-2}$, and $\int \tau_{21} v\,\mathrm{d}v$ is the rest-frame velocity integrated optical depth. In the case of PKS\,1657$-$298, $N_{\mathrm{H,X}}$ = 0.6~$\times$~10$^{22}$\,cm$^{-2}$ and $\int \tau_{21} v\,\mathrm{d}v$ = 2.9\,km\,s$^{-1}$. Thus we find an upper limit for spin temperature of $T_s \lesssim$ 1130 $\pm$ 380\,K, assuming a covering fraction of 1 based on the compactness seen on VLBI scales. Given that most H{\sc i} absorbers have been found to have measured spin temperatures in the range $10^2$--$10^4$\,K \citep{Curran:2012hj,Kanekar:2014kj} as well as the predictions of higher spin temperatures near the central region of an AGN compared to standard Galactic spin temperatures \citep{Bahcall:1969kq}, this result is entirely consistent with previous work. 

Aside from the upper limit above (which is based on the assumption that $N_{\mathrm{H,X}}$ = $N_{\mathrm{HI}}$), there are other factors to consider that will influence these measurements. Firstly, since $N_{\mathrm{H,X}}$ is a measure of the total hydrogen column, it will reflect the presence of neutral, ionised and molecular hydrogen while $N_{\mathrm{HI}}$ only probes the neutral hydrogen along the line of sight. So we would in all cases expect $N_{\mathrm{H,X}}$ $\ge$ $N_{\mathrm{HI}}$, but the relative fractions of ionised, neutral and molecular within the total $N_{\mathrm{H,X}}$ would vary depending on the properties of the AGN and local interstellar medium. Given the results found by \citet{Wilson:1998it} of a low ionisation fraction on the order of 10$^{-2}$ to 10$^{-5}$ in NGC\,2639, it seems unlikely that the combination of ionised and molecular hydrogen could account for 90\% of the total hydrogen column. However, the AGN X-ray luminosity is three orders of magnitude brighter in the case of PKS\,1657$-$298 than in NGC\,2639 and thus it also seems extremely unlikely that the entire hydrogen column is neutral along the line of sight to the AGN. It is worth noting that for kinetic temperatures $T_k < 1000$\,K in the cold neutral medium (CNM), $T_s$ is basically equal to $T_k$ \citep{Liszt:2001bn,Roy:2006el}, although the gas we detect toward the AGN may not be in the CNM phase due to its likely proximity to the central black hole. In any case, the true neutral fraction making up $N_{\mathrm{H,X}}$ (likely located somewhere between 10--90\%) will thus lead to a lower spin temperature than the upper limit of $\sim$1100\,K. 

We can consider the kinetic temperature implied by the line-width of our absorption line profile, as a means of comparing it to what we see in the spin temperature. We derive the kinetic temperature as in \citet{2012MNRAS.421.3159M}, namely

\begin{equation}
T_k \le \frac{m_H \sigma_v^2}{k},
\end{equation}
~\\
where $m_H$ is the mass of the hydrogen atom in kg, $\sigma_v$ is the velocity dispersion of the line and $k$ is the Boltzmann constant. Our measured FWHM line-width for PKS\,1657$-$298 is 63\,km\,s$^{-1}$, which corresponds to a $\sigma_v$ of 27\,km\,s$^{-1}$. This gives an upper limit estimate for $T_k$ of $8.6 \times 10^4$\,K, however we note that this does not take into account the effects of rotation or turbulent broadening both of which will result in a non-thermally broadened line. It is possible that including these effects could reduce the width of the line by the factor of $\sim$60 required to match the spin temperature upper limit.

It is also important to note the assumption made above that the H{\sc i} and X-rays are spatially coincident. There is no guarantee based on these data that we are probing the same physical regions that we see in H{\sc i} absorption and in X-ray absorption. While we might expect the X-ray emission to arise from a reasonably uniform area surrounding the torus of the AGN, it remains unclear whether the H{\sc i} absorption we see arises from a similarly uniform region or if it instead comes from a clumpy region and we are detecting cold clouds along the line of sight (e.g. as in the case of \citealt{Maccagni:2014kq}). However, our results here showing that H{\sc i} absorbers are more likely to feature significant soft X-ray absorption support the suggestion of co-spatiality, alongside the results presented by \citet{Ostorero:2010eb} of a tentative but statistically robust positive correlation between $N_{\mathrm{H,X}}$ and $N_{\mathrm{HI}}$. Further data in this area, as well as detailed information on more comparable spatial scales (i.e. through VLBI studies of the H{\sc i} absorption distribution in associated systems), will be key in conclusively answering this question of spatial coincidence.

\section{Conclusion}\label{disc}
We have presented here analysis of the radio and X-ray properties of 94 galaxies searched for H{\sc i} absorption, including 20 absorbers and 74 non-detections. By analysing their X-ray properties, we find that absorbers are more likely to feature an absorbed X-ray spectrum. We also find a statistically-significant trend at 4.71$\sigma$ using survival analysis indicating correlation between H{\sc i} optical depth and X-ray hardness, as well as a higher proportion of absorbers with increasing X-ray hardness. These results are all supportive of correlation between H{\sc i} absorption and soft X-ray absorption, and are strongly suggestive that these absorption mechanisms are tracing similarly-distributed populations of gas. In future investigations, we plan to further explore this correlation in order to confirm the spatial distribution of $N_{\rm HI}$ and $N_{\rm H}$, and determine whether we are detecting absorption from inside the sphere of influence of the black hole (probing a torus or wind), from a few 10s of pc in the galactic nucleus (probing the galactic inflow and of the gas available for BH growth), or a few kpc in the whole galaxy (probing the large-scale galaxy evolution). More detailed analysis and complex modelling of high signal-to-noise X-ray spectra where available, in comparison with studies in the literature \citep[e.g.][]{Buchner:2017fw}, will aid in separating galactic-scale from nuclear-scale absorption contributions.

We have also presented the results of a pilot study conducted with the commissioning telescope BETA, as part of preparations for FLASH science with ASKAP. Our sample of five X-ray absorbed galaxies revealed a new H{\sc i} absorption detection at a redshift of $z$ = 0.42, towards PKS\,1657$-$298, which also showed the most X-ray absorption in its 3XMM-DR4 catalogue spectrum. PKS\,1657$-$298 is detected across the electromagnetic spectrum, particularly at mid-infrared wavelengths as seen in WISE data. Although we do not currently have a optical spectroscopic confirmation of the radio source redshift, the $K$-$z$ relation based on analysis by \citet{Willott:2003ep} is consistent with the redshift as measured in H{\sc i}. In analysing the properties of the H{\sc i} absorption towards PKS\,1657$-$298, we find values of optical depth $\tau$ = 0.046, effective line-width, $\Delta v_{\rm eff}$ = 63\,km\,s$^{-1}$ and $N_{\rm HI}$ = 5.3 $\times$ 10$^{20}$\,cm$^{-2}$, assuming a spin temperature of 100\,K and a covering fraction of 1.

Our analysis of archival {\it XMM-Newton} and {\it Chandra} data for PKS\,1657$-$298 reveals consistent results for intrinsic X-ray luminosity $L_x \sim 5 \times 10^{44}$\,erg\,s$^{-1}$ and an estimated hydrogen column density in the galaxy of $N_{\mathrm{H,X}} \sim 6 \times 10^{21}$\,cm$^{-2}$. In the {\it Chandra} data, a simple absorbed power-law model is not sufficient to take into account high-energy residuals, which are instead well-fit by assuming the presence of both a narrow 6.4\,keV Fe line and a 7.1\,keV Fe K-shell absorption edge both at a redshift of $z$ = 0.42. The {\it XMM-Newton} data does not have high enough signal-to-noise to confirm this, but the data are statistically consistent with the results from {\it Chandra}. We conclude that this is the most likely scenario for interpreting the high-energy components we see in our averaged spectra, however we note that we cannot currently distinguish these from a P-Cygni profile signifying an outflow in the host galaxy. The combination of X-ray absorption with H{\sc i} absorption allows us to place an upper limit on the spin temperature of gas in PKS\,1657$-$298 of $T_s \lesssim$ 1130 $\pm$ 380\,K.

Our X-ray pilot sample has shown the potential for expanding this line of investigation with ASKAP Early Science and beyond, as well as with telescopes around the world able to explore neutral gas at these intermediate redshifts. The BETA sample was extremely limited in sensitivity to sources above 1\,Jy, however with full ASKAP we will have access to the radio source population down to $\sim$50\,mJy which opens up an entire new window on the parameter space we can explore. There is particularly strong synergy between the H{\sc i} absorption population to be discovered with FLASH and the all-sky survey capability in X-rays of {\it eROSITA} \citep{Merloni:2012ug}, where we expect to greatly expand on the results described here. It is also critical to further explore the question of spatial correlation between X-rays and radio emission near the core of radio galaxies, which will be made possible in the southern hemisphere with extension of VLBI capability to lower frequencies using baselines between ASKAP and the 0.7--4\,GHz wide-band receiver of the Parkes telescope \citep{Manchester:2015uu}. In the more distant future, the powerful combination of the Square Kilometre Array \citep{Dewdney:2009bq} with {\it Athena} \citep{Barret:2013uba} will enable studies of these objects at sensitivities that are orders of magnitude beyond what we can currently achieve with existing telescopes, up to $z$ = 3 and beyond across a much wider range of galaxy types, luminosities, and evolutionary states.

\section*{Acknowledgments} 
We would like to thank our referee for useful comments to improve the clarity and quality of this paper. We also thank R.D.\,Ekers, J.\,Buchner, M.\,Salvato and G.A.\,Rees for useful discussions regarding this work. This research was conducted as part of the Australian Research Council Centre of Excellence for All-sky Astrophysics (CAASTRO), through project number CE110001020. We acknowledge and thank the ASKAP commissioning team for the use of BETA when time was available, thus enabling the studies described in this paper. The Australian SKA Pathfinder is part of the Australia Telescope National Facility which is managed by CSIRO. Operation of ASKAP is funded by the Australian Government with support from the National Collaborative Research Infrastructure Strategy. ASKAP uses the resources of the Pawsey Supercomputing Centre. Establishment of ASKAP, the Murchison Radio-astronomy Observatory and the Pawsey Supercomputing Centre are initiatives of the Australian Government, with support from the Government of Western Australia and the Science and Industry Endowment Fund. We acknowledge the Wajarri Yamatji people as the traditional owners of the Observatory site.

We acknowledge use of {\sc asurv} Rev 1.2 \citep{Lavalley:1992wf}, which implements the methods presented in \citet{Isobe:1986iu}. This research has made use of the VizieR catalogue access tool, CDS, Strasbourg, France, NASA's Astrophysics Data System, and the NASA/IPAC Extragalactic Database (NED) which is operated by the Jet Propulsion Laboratory, California Institute of Technology, under contract with the National Aeronautics and Space Administration. We also acknowledge use of APLpy, an open-source plotting package for Python hosted at http://aplpy.github.com, and Astropy, a community-developed core Python package for Astronomy \citep{Collaboration:2013cd}.

This publication makes use of data products from: 1) the Wide-field Infrared Survey Explorer\citep{Wright:2010in}, which is a joint project of the University of California, Los Angeles, and the Jet Propulsion Laboratory/California Institute of Technology, funded by the National Aeronautics and Space Administration, 2) observations obtained with {\it XMM-Newton}, an ESA science mission with instruments and contributions directly funded by ESA Member States and NASA, 3) the {\it Chandra} Data Archive and the {\it Chandra} Source Catalog, and software provided by the {\it Chandra} X-ray Center (CXC) in the application packages CIAO, ChIPS, and Sherpa, 4) the Two Micron All Sky Survey, which is a joint project of the University of Massachusetts and the Infrared Processing and Analysis Center/California Institute of Technology, funded by the National Aeronautics and Space Administration and the National Science Foundation, and 5) the Digitized Sky Surveys, which were produced at the Space Telescope Science Institute under U.S. Government grant NAG W-2166. The images of these surveys are based on photographic data obtained using the Oschin Schmidt Telescope on Palomar Mountain and the UK Schmidt Telescope. The plates were processed into the present compressed digital form with the permission of these institutions.


\end{document}